%% file: main.tex
\documentclass[onefignum,onetabnum]{siamart171218}

\usepackage{lipsum}
\usepackage{amsfonts}
\usepackage{graphicx}
\usepackage{epstopdf}
\usepackage{algorithmic}

\usepackage{amssymb}
\usepackage{amsmath}

\usepackage{comment}
\usepackage{graphicx}
\usepackage{caption}
\usepackage{subcaption}
\usepackage{float}
\usepackage{xcolor}
\usepackage{booktabs}
\usepackage{siunitx}
\input{macros}

\let\citep\cite
\providecommand{\sep}{\unskip, }
\ifpdf
  \DeclareGraphicsExtensions{.eps,.pdf,.png,.jpg}
\else
  \DeclareGraphicsExtensions{.eps}
\fi

\newsiamremark{remark}{Remark}
\newsiamremark{hypothesis}{Hypothesis}
\crefname{hypothesis}{Hypothesis}{Hypotheses}
\newsiamthm{claim}{Claim}

\headers{A finite element method for fluctuating Navier--Stokes equation}
{D. Gourzoulidis, M. Gallo, S. Elkantassi, T. Kay, and S. Kalliadasis}
\title{A finite element method for fluctuating Navier--Stokes equations%
\thanks{\funding{
Dimitrios Gourzoulidis, Toby Kay, and Serafim Kalliadasis acknowledge financial support from the ERC--EPSRC Frontier Research Guarantee through Grant No.~588 EP/X038645, the ERC through Advanced Grant No.~247031, and the EPSRC through Grant Nos.~EP/L025159 and EP/L020564.
Soumaya Elkantassi acknowledges financial support from the Air Force Office of Scientific Research under award number FA8655-24-1-7009.
Mirko Gallo acknowledges financial support from the European Union. Views and opinions expressed are, however, those of the authors only and do not necessarily reflect those of the European Union or the European Research Council Executive Agency. Neither the European Union nor the granting authority can be held responsible for them. This work is supported by an ERC grant, ERC-STG E-Nucl., Grant Agreement ID: 101163330.
}}}

\author{
Dimitrios Gourzoulidis\thanks{Department of Chemical Engineering, Imperial College London, London SW7 2AZ, United Kingdom
  (\email{d.gourzoulidis@imperial.ac.uk}, \email{t.kay@imperial.ac.uk}, \email{s.kalliadasis@imperial.ac.uk}).}
\and Mirko Gallo\thanks{Department of Mechanical and Aerospace Engineering, Sapienza University of Rome, Rome 00184, Italy
  (\email{mirko.gallo@uniroma1.it}).}
\and Soumaya Elkantassi\thanks{Department of Operations, University of Lausanne, Lausanne 1015, Switzerland
  (\email{soumaya.elkantassi@unil.ch}).}
\and Toby Kay\footnotemark[2]
\and Serafim Kalliadasis\footnotemark[2]
}
\usepackage{amsopn}

\makeatletter
\newcommand*{\addFileDependency}[1]{
  \typeout{(#1)}
  \@addtofilelist{#1}
  \IfFileExists{#1}{}{\typeout{No file #1.}}
}
\makeatother

\ifpdf
\hypersetup{
  pdftitle={A finite element method for fluctuating Navier-Stokes equation},
    pdfauthor={Dimitrios Gourzoulidis, Mirko Gallo, Soumaya Elkantassi, Toby Kay, and Serafim Kalliadasis}
}
\fi

\begin{document}

\maketitle

\begin{abstract}
We introduce a finite-element framework for simulating thermal fluctuations in compressible fluids governed by the fluctuating Navier-Stokes equations. The method is designed to preserve the fundamental fluctuation-dissipation balance at the discrete level. This is achieved by defining the stochastic forcing term in the weak formulation, ensuring its covariance is proportional to the discrete viscous dissipation operator. A nodal quadrature rule is employed to eliminate unphysical mesh-scale correlations. The time integration is performed using the Crank-Nicolson scheme to maintain numerical stability and accuracy. The proposed approach is numerically validated in one, two, and three spatial dimensions, demonstrating its capability to correctly capture equilibrium fluctuation statistics across various discretisation parameters.
\end{abstract}

\begin{keywords}
Finite--element methods \sep Stochastic Navier--Stokes equations \sep Stochastic differential equations \sep Fluctuation--dissipation balance
\end{keywords}

\begin{AMS}
 65M60, 60H15, 76M10, 76D05, 35Q30
\end{AMS}
\section{Introduction}

Thermal fluctuations play a critical role in a vast range of physical phenomena, acting as the driving force behind Brownian motion at microscopic scales and, at larger scales, as the trigger for activated processes such as cavitation and boiling \cite{Gallo_Magaletti_Casciola_2021, gallo2023nanoscale,gallo2022thermal,gallo2024vapor, gallo2025complex}, condensation \cite{teodori2025nanodroplet} crystal formation \cite{Lutsko2019}, and hydrodynamic instabilities  \cite{barker2023fluctuating}. 
While thermal noise is often assumed to be negligible at macroscopic scales, where deterministic dynamics are expected to dominate, recent studies have shown that the interaction between slow macroscopic modes and fast microscopic fluctuations is far from trivial. In particular, the coupling between thermal fluctuations and turbulence at scales approaching the Kolmogorov length can lead to non-negligible effects \cite{bandak2022dissipation, bell2022thermal}. Moreover, giant non-equilibrium fluctuations in diffusion-driven systems have been experimentally observed \cite{vailati1997giant}, for instance in light-scattering and shadowgraph measurements of fluids subjected to concentration or temperature gradients, where fluctuations are amplified over macroscopic length scales. These phenomena were subsequently framed and rationalised from both theoretical and numerical perspectives within the context of fluctuating hydrodynamics and non-equilibrium statistical mechanics \cite{vailati1998nonequilibrium, donev2014reversible, eyink2024kraichnan, bussoletti2025emergence, bussoletti2026non}. 

These results demonstrate that even at scales traditionally considered deterministic, the interplay between fluctuations and nonlinear hydrodynamics may qualitatively alter the system behaviour. It has also been argued that thermal noise alone may trigger spontaneous stochasticity in high–Reynolds-number turbulence, with important consequences for predictive modelling and engineering of fluid systems \cite{bandak2024spontaneous}. These are some of the reason why continuum-based computational models are augmented  with thermal fluctuations. This is achieved through the framework of fluctuating hydrodynamics, which extends the classical Navier–Stokes equations by introducing a stochastic stress tensor whose statistical properties are constrained by the fluctuation–dissipation theorem \cite{LandauLifshitz1980StatPhys}. 
Nowadays, there is a growing interest in numerical schemes specifically designed to accurately discretise stochastic equations of the fluctuating-hydrodynamics type \cite{donev2010accuracy, balboa2012staggered, MagalettiGalloPerezCarrilloKalliadasis}.  

A central pillar of fluctuating hydrodynamics is the fluctuation–dissipation theorem, which prescribes the statistical structure of the stochastic stress tensor. It ties the stochastic forcing coefficient and its covariance to viscous dissipation, so that the system attains the correct statistical equilibrium and reproduces the expected fluctuation spectra. The primary challenge for computational methods is therefore to discretise the governing equations while preserving this fundamental balance at the discrete level \cite{donev2010accuracy}. If the fluctuation–dissipation condition is violated numerically, the method generates unphysical artifacts such as excessive damping of fluctuations or artificial energy growth, distorting the equilibrium moments.



Several numerical strategies have been developed to compute numerical solutions of the equations of fluctuating hydrodynamics. Finite-difference methods \cite{MagalettiGalloPerezCarrilloKalliadasis,Florencioetatll2012} and finite-volume schemes \cite{DonevNonakaSunFaiGarciaBell2014_LowMach_FH_Mixing,DonevVandenEijndenGarciaBell2010_FV_FH_Accuracy} have been widely applied. These methods often provide excellent discrete conservation properties and algorithmic simplicity. However, achieving high-order accuracy on complex, unstructured geometries can be challenging. As a result, the finite element method is a powerful alternative, offering natural handling of complex geometries and a robust variational framework for discretisation.

Despite these strengths, the application of finite element methods to stochastic hydrodynamics has been limited. A significant challenge is that standard Galerkin discretisations introduce artificial, non-physical correlations in the fluctuation fields. This artifact arises from the global support of the basis functions, which leads to a non-diagonal mass matrix and distorts the local equilibrium statistics derived theoretically. Structure-preserving discretisations are developed for simpler thermally forced SPDEs \cite{MartinezLeraDeCorato2024,EspanolDonev2015_NanoparticleIsothermalFH_CoarseGraining}. Moreover, structure-preserving finite-element discretisations and convergence analyses are available for incompressible stochastic Navier–Stokes-type formulations and related mixed methods \cite{DECORATO2016632,LI2021398,Darryl2015,FengQiu2021_StochasticStokes_MixedFEM}. Extending finite-element techniques to the fully coupled, nonlinear Landau–Lifshitz fluctuating Navier–Stokes equations remains an active research direction. A key difficulty is preserving discrete fluctuation–dissipation consistency on general meshes under realistic boundary conditions.

In this work, we introduce a novel finite-element methodology for the isothermal, compressible fluctuating Navier-Stokes equations that directly addresses this challenge. Our approach is structure-preserving by design: we construct a discrete stochastic forcing whose covariance is proportional to the discrete viscous dissipation operator, ensuring the fluctuation-dissipation balance is maintained at the discrete level. We use a nodal quadrature rule for the key inner products, which yields diagonal mass matrices and ensures that fluctuations are local and physical. The resulting scheme is integrated in time using a $\theta$-scheme, with the Crank-Nicolson method. 

The remainder of this paper is organized as follows. In Section~\ref{FHM}, we present the  fluctuating hydrodynamics equations. 
Section~\ref{Variational_formulation} describes the variational formulation and details the full numerical discretisation, including the Galerkin projection, the discrete noise, and the temporal integration scheme. Section~\ref{sec:linearization} analyzes the linearised equations, demonstrating how our discretisation preserves the fluctuation-dissipation balance for the Crank-Nicolson scheme. Section~\ref{sec:numerical_results} provides comprehensive numerical validation in one, two, and three spatial dimensions, confirming the method's accuracy and robustness. We conclude in Section 6 with a summary of findings and a discussion of potential future extensions.


\section{Fluctuating hydrodynamics model}\label{FHM}
We model thermal fluctuations in a compressible viscous fluid using the Landau--Lifshitz fluctuating hydrodynamics framework to capture spontaneous fluctuations that are consistent with statistical mechanics at thermodynamic equilibrium. In this setting, the macroscopic balance laws are complemented by a stochastic stress tensor $\widetilde{\boldsymbol{\tau}}$ that represents the effect of thermal fluctuations at the microscale and is modeled as a Gaussian random field. The resulting random forcing is not arbitrary: its covariance is tied to the viscous dissipation so that the dynamics satisfies a fluctuation--dissipation principle and reproduces the correct equilibrium fluctuation levels. Preserving this structure under discretisation is central to our approach, and it will guide the variational formulation and the numerical scheme developed in the next section.

We consider a compressible fluid satisfying
\begin{align}
\dfrac{\partial \rho}{\partial t}+\nabla\!\cdot(\rho\,\mathbf u) &= 0,\label{eq_NS1}\\
\dfrac{\partial (\rho\,\mathbf u)}{\partial t}
+\nabla\!\cdot(\rho\,\mathbf u\otimes\mathbf u)
&= \nabla\!\cdot\!\Big[
 -\,p\,\mathbf I
 + \mu\,\big(\nabla\mathbf u+(\nabla\mathbf u)^{\!\top}\big)
 - \frac{2}{d}\,\big(\nabla\!\cdot(\mu\,\mathbf u)\big)\mathbf I
\Big]
+ \nabla\!\cdot\widetilde{\boldsymbol{\tau}},
\label{eq_NS2}
\end{align}
where $t\in[0,\infty)$, $\rho:\mathbb{R}^d\times[0,\infty)\to\mathbb{R}$ is the density,
$p:\mathbb{R}^d\times[0,\infty)\to\mathbb{R}$ is the pressure, and
$\mathbf u:\mathbb{R}^d\times[0,\infty)\to\mathbb{R}^d$ is the velocity field.
The tensor $\mathbf I$ denotes the identity, and $\mu$ is the (dynamic) viscosity coefficient.
The stochastic stress tensor $\widetilde{\boldsymbol{\tau}}=\{\widetilde{\tau}_{ij}\}_{i,j=1}^d$
is assumed to be mean-zero and delta-correlated in space and time 
\begin{align*}
\langle \widetilde{\tau}_{ij}(\mathbf x,t) \rangle &= 0,\\
\langle \widetilde{\tau}_{ij}(\mathbf x,t)\,\widetilde{\tau}_{pq}(\mathbf y,\tau)\rangle
&= \big(\delta_{ip}\delta_{jq}+\delta_{iq}\delta_{jp}-\tfrac{2}{d}\delta_{ij}\delta_{pq}\big)\,
2\,\mu\,k_B\,T\,\delta(\mathbf x-\mathbf y)\,\delta(t-\tau),
\end{align*}
where $T>0$ is the temperature and $\delta_{ij}$ denotes the Kronecker delta. This structure ensures the fluctuation-dissipation balance is satisfied for the system \cite{LandauLifshitz1980StatPhys, fox1970contributions}. Here $\langle\cdot\rangle$ denotes expectation with respect to the underlying probability space. It is important to stress that the equations of fluctuating hydrodynamics, when formally written as a system of stochastic partial differential equations, are not mathematically well defined, despite the fact that they are often treated in this way in the physics literature. In a rigorous sense, they must instead be regarded as an effective theory valid at small wavenumbers \cite{glorioso2018lectures}. Accordingly, the continuum limit is not well posed in the usual sense: derivatives are meaningful only in the presence of an implicit ultraviolet cutoff, and differential operators must ultimately be understood as discrete.

Section~\ref{Variational_formulation} develops the variational formulation and the numerical discretisation of \eqref{eq_NS1}--\eqref{eq_NS2}. We first derive the weak form in Section~\ref{weak_formulation}. We then introduce the finite element mesh and discrete spaces and obtain the semi-discrete formulation in Section~\ref{Semi-discrete formulation}. Finally, after discretizing the time, we arrive at the fully discrete scheme in Section~\ref{full_discretisation}.
\section{Variational formulation and numerical discretisation}\label{Variational_formulation} 
Our goal is a finite element discretisation that mirrors the continuum fluctuation--dissipation structure. We therefore start from a weak formulation and define the discrete stochastic forcing using the same finite element spaces and weak operators. Related approaches have been proposed for the stochastic linear and non linear diffusion equations (see, e.g., \cite{Torre2015,MartinezLeraDeCorato2024}) and for fluctuating Navier--Stokes models (see, e.g., \cite{MartinezLeraDeCorato2024}).

\subsection{Weak formulation} \label{weak_formulation}
A weak formulation is a rewritten version of the governing partial differential equations, also called the strong form, that is defined through integral identities against test functions, rather than requiring the equations to hold pointwise. In practice, one multiplies the strong form by sufficiently regular test functions, integrates over the spatial domain, and applies integration by parts to transfer derivatives from the solution onto the test functions; see \cite{yan2005galerkin,LordPowellShardlow2014}. For simplicity, we assume boundary conditions under which the boundary terms produced by integration by parts vanish, for example periodic and homogeneous natural boundary conditions; the specific boundary conditions used in the numerical experiments are stated later.

Let $\Omega\subset\mathbb{R}^d$ be a bounded domain and $T_f>0$.
We introduce the spaces
$$V := [H^1(\Omega)]^d, \qquad W := H^1(\Omega),$$
and seek $\rho:(0,T_f)\to W$ and $\mathbf u:(0,T_f)\to V$ satisfying, for almost every $t\in(0,T_f)$,
the weak formulation of \eqref{eq_NS1}-\eqref{eq_NS2}, i.e., find $(\rho(t),\mathbf u(t))\in W\times V$ such that
\begin{align}
(\partial_t \rho,\psi) - \int_\Omega \rho\,\mathbf u\cdot\nabla\psi\,\mathrm dx
&= 0, \qquad \forall \psi\in W, \label{eq:weak-continuity}\\
(\partial_t(\rho\mathbf u),\mathbf v)
- \int_\Omega \rho\,\mathbf u\otimes\mathbf u : \nabla\mathbf v\,\mathrm dx
+ \int_\Omega \mu\big(\nabla\mathbf u+(\nabla\mathbf u)^{\!\top}\big):\nabla\mathbf v\,\mathrm dx
&\nonumber\\
\quad - \dfrac{2}{d}\int_\Omega \big(\nabla\cdot(\mu\mathbf u)\big)\,(\nabla\cdot\mathbf v)\,\mathrm dx
- T\int_\Omega \rho\,\nabla\cdot\mathbf v\,\mathrm dx
+ \int_\Omega \widetilde{\boldsymbol\tau}:\nabla\mathbf v\,\mathrm dx
&= 0, \qquad \forall \mathbf v\in V. \label{eq:weak-momentum}
\end{align}
where $(\cdot,\cdot)$ denotes the $L^2(\Omega)$ inner product and
$\mathbf A:\mathbf B = \sum_{i,j=1}^d A_{ij}B_{ij}$. 
Equations \eqref{eq:weak-continuity}--\eqref{eq:weak-momentum} follow by multiplying
\eqref{eq_NS1}--\eqref{eq_NS2} by test functions and integrating by parts in space.
In deriving \eqref{eq:weak-momentum}, we use the ideal-gas equation of state
$p=R\rho T$. We work with nondimensional variables and choose units such that
the specific gas constant satisfies $R=1$; hence, $p=\rho T$.


We define the fluctuating stress functional in the weak formulation via
\begin{equation}\label{eq:def-F}
  F(\mathbf{v},t)
:= \int_\Omega \left(\nabla\!\cdot \widetilde{\boldsymbol{\tau}}(x,t)\right)
   \cdot \mathbf{v}\,dx
= -\int_\Omega \widetilde{\boldsymbol{\tau}}(x,t):\nabla\mathbf{v}\,dx,
\quad \mathbf{v}\in V.  
\end{equation}
where the second identity follows by integration by parts under the assumed boundary conditions.
We also introduce the viscous bilinear form 
\begin{equation}
\label{eq:viscous-bilinear-form}
a(\mathbf{v},\mathbf{w})
:= \int_\Omega \mu\big(\nabla\mathbf{v}+(\nabla\mathbf{v})^{\top}\big)
                 :\nabla\mathbf{w}\,dx
   - \dfrac{2}{d}\int_\Omega
     \big(\nabla\!\cdot(\mu\mathbf{v})\big)\,
     \big(\nabla\!\cdot\mathbf{w}\big)\,dx,
\quad \mathbf{v},\mathbf{w}\in V,
\end{equation}
which is the sum of the third and fourth term in \eqref{eq:weak-momentum}.
Assuming the standard Landau--Lifshitz statistics for $\widetilde{\boldsymbol{\tau}}$,
the resulting forcing functional $F(\cdot,t)$ is mean-zero and has covariance
\begin{equation}
\label{eq:weak-covariance}
\langle F(\mathbf v,t)\,F(\mathbf w,\tau)\rangle
= 2\,k_B T\,a(\mathbf v,\mathbf w)\,\delta(t-\tau).
\end{equation}
The derivation is detailed in~\ref{Appendix A}.

Equation \eqref{eq:weak-covariance} expresses the fluctuation--dissipation principle in variational form: the same bilinear form $a(\cdot,\cdot)$ that represents viscous dissipation in \eqref{eq:weak-momentum} also determines the covariance structure of the thermal forcing. As a result, the stochastic forcing is consistent with equilibrium statistical mechanics and the
formulation reproduces the correct equilibrium fluctuation levels. In particular, this matching avoids unphysical energy injection or artificial damping caused by an inconsistent noise model. In Section~\ref{sec:linearization} and \ref{sec:numerical_results}, we show that the fully discrete scheme preserves this balance, so that thermal fluctuations are neither amplified nor suppressed by the numerical discretisation of the viscous term.


\subsection{Semi-discrete finite element formulation}\label{Semi-discrete formulation}
We now introduce the spatial discretisation used in the semi-discrete scheme. Let $\Omega\subset\mathbb{R}^d$ be a bounded polygonal domain (polyhedral if $d=3$). For $h>0$, let $\mathcal{T}_h$ be a conforming triangulation of $\Omega$ into triangles ($d=2$)
or tetrahedra ($d=3$), and define the mesh size by $h := \max_{K\in\mathcal{T}_h} \operatorname{diam}(K)$. We denote by $\mathcal{N}_h$ the set of mesh vertices and by $\mathcal{N}_h^{\mathrm{int}}\subset \mathcal{N}_h$ the subset of interior nodes. 

On $\mathcal{T}_h$ we use continuous, piecewise affine finite elements. The associated scalar space is
\begin{equation*}
V_h
:= \Big\{ v_h \in C^0(\overline{\Omega}) \;:\;
        v_h|_K \in \mathbb{P}_1(K)\ \ \forall\,K\in\mathcal{T}_h \Big\},
\end{equation*}
where $\mathbb{P}_1(K)$ denotes the polynomials of total degree at most one on $K$. 

We obtain the semi-discrete scheme by applying a Galerkin approximation in space \cite{Ciarlet2002}. Using the scalar space $V_h$ defined above, we set
$$
W_h := V_h, \qquad \mathbf V_h := [V_h]^d,
$$
and seek $(\rho_h,\mathbf u_h)\in W_h\times \mathbf V_h$ such that
\eqref{eq:weak-continuity}--\eqref{eq:weak-momentum} hold for all test functions in
$W_h$ and $\mathbf V_h$.
We define the discrete inner product $(\phi,\psi)_{0,h}$ as a mass-lumped approximation of the $L^2(\Omega)$ inner product obtained via the trapezoidal quadrature rule. For continuous, piecewise linear finite elements this reduces to the nodal formula
\begin{equation}\label{eq:discrete_inner_product}
(\phi,\psi)_{0,h} := \sum_{x_i\in\mathcal N_h} \omega_i\,\phi(x_i)\,\psi(x_i),
\end{equation}
where $\omega_i>0$ are the associated quadrature weights, equivalently, the diagonal entries of the lumped mass matrix. This choice simplifies the computation of the stochastic forcing and, as shown in Section~\ref{sec:linearization}, yields a discrete equilibrium covariance that is diagonal at the nodes.
We now use the vertex-based inner product $(\cdot,\cdot)_{0,h}$ on
scalar, vector, and tensor fields, and define the discrete viscous
bilinear form
\begin{equation*}
a_h(\mathbf{v}_h,\mathbf{w}_h)
:= \big(\mu\big(\nabla\mathbf{v}_h+(\nabla\mathbf{v}_h)^{\top}\big),
        \nabla\mathbf{w}_h\big)_{0,h}
   - \dfrac{2}{d}\big(\nabla\!\cdot(\mu\mathbf{v}_h),
                     \nabla\!\cdot\mathbf{w}_h\big)_{0,h},
\quad \mathbf{v}_h,\mathbf{w}_h\in\mathbf{V}_h,
\end{equation*}
 where $a_h(\cdot, \cdot)$ is the discrete approximation of the viscous bilinear form $a(\cdot, \cdot)$.
We further introduce a discrete fluctuating force functional
$F_h(\cdot,t):\mathbf{V}_h\to\mathbb{R}$ with covariance
\begin{equation}
\label{eq:weak-covariance-discrete}
\left\langle F_h\left(\mathbf{v}_h,t\right)\,F_h\left(\mathbf{w}_h,\tau\right)\right\rangle
= 2\,k_B T\,a_h(\mathbf{v}_h,\mathbf{w}_h)\,\delta(t-\tau),
\quad \forall\,\mathbf{v}_h,\mathbf{w}_h\in\mathbf{V}_h.
\end{equation}
The semi-discrete problem reads: find
$$
\rho_h : (0,T_f)\to W_h, \qquad
\mathbf{u}_h : (0,T_f)\to \mathbf{V}_h,
$$
such that, for almost every $t\in(0,T_f)$,
\begin{alignat}{2}
(\partial_t \rho_h(t), \psi_h)_{0,h}
- (\rho_h(t)\mathbf u_h(t), \nabla \psi_h)_{0,h}
&= 0,
&\qquad& \forall\,\psi_h \in W_h,
\label{eq:disc-cont}\\
(\partial_t (\rho_h(t)\mathbf u_h(t)), \mathbf v_h)_{0,h}
- (\rho_h(t)\mathbf u_h(t)\otimes \mathbf u_h(t), \nabla\mathbf v_h)_{0,h}
& & & 
\notag\\
\qquad {}+ a_h(\mathbf u_h(t),\mathbf v_h)
- T\,(\rho_h(t), \nabla\!\cdot \mathbf v_h)_{0,h}
- F_h(\mathbf v_h,t)
&= 0,
&\qquad& \forall\,\mathbf v_h \in \mathbf V_h,
\label{eq:disc-mom}
\end{alignat}
Equation \eqref{eq:weak-covariance-discrete} is the spatially discrete analogue of
\eqref{eq:weak-covariance}: the same bilinear form $a_h(\cdot,\cdot)$ that represents viscous
dissipation in \eqref{eq:disc-mom} also determines the covariance structure of the discrete thermal forcing. As a result, the semi-discrete stochastic forcing is consistent with equilibrium statistical mechanics and yields the correct equilibrium fluctuation levels at the level of the spatial discretisation. 
\subsection{Time discretisation}\label{full_discretisation}
We now discretise the semi-discrete system \eqref{eq:disc-cont}--\eqref{eq:disc-mom} in time using a semi-implicit $\theta$--scheme for the deterministic part and an Euler–Maruyama approximation for the stochastic perturbation. We use a linearized decoupling in which the continuity equation is advanced semi-implicitly in $\rho$ using the explicit velocity $\mathbf u_h^{n}$, while the momentum equation is advanced using the explicit density $\rho_h^{n}$. Let $0=t_0<t_1<\dots<t_N=T_f$ be a uniform grid with $\Delta t:=t_{n+1}-t_n$, and denote
$$\rho_h^n \approx \rho_h(t_n), \qquad
\mathbf{u}_h^n \approx \mathbf{u}_h(t_n).
$$
For $0\le \theta \le 1$, we define the convex combinations
$$
\rho_h^{n,\theta} := \theta \rho_h^{n+1} + (1-\theta)\rho_h^{n},\qquad
\mathbf{u}_h^{n,\theta} := \theta \mathbf{u}_h^{n+1} + (1-\theta)\mathbf{u}_h^{n}.
$$
Given $(\rho_h^n,\mathbf u_h^n)\in W_h\times\mathbf V_h$, we compute
$(\rho_h^{n+1},\mathbf u_h^{n+1})\in W_h\times\mathbf V_h$ from
\begin{equation} \label{eq:theta-scheme}
\begin{aligned}
\bigg(\dfrac{\rho_h^{n+1}-\rho_h^{n}}{\Delta t}, \psi_h\bigg)_{0,h}
- \big(\rho_h^{n,\theta}\mathbf{u}_h^{n}, \nabla \psi_h\big)_{0,h}
&= 0,
\qquad \forall\,\psi_h\in W_h,\\[0.4em]
\bigg(\dfrac{\rho_h^{n}\mathbf{u}_h^{n+1}-\rho_h^{n}\mathbf{u}_h^{n}}{\Delta t},
      \mathbf{v}_h\bigg)_{0,h}
+ a_h(\mathbf{u}_h^{n,\theta},\mathbf{v}_h)
- T\,(\rho_h^{n}, \nabla\!\cdot \mathbf v_h)_{0,h}
&\\
\quad
- (\rho_h^{n}\mathbf u_h^{n,\theta}\otimes \mathbf u_h^{n}, \nabla\mathbf v_h)_{0,h}
- F_h^{n}(\mathbf v_h)
&= 0,
\qquad \forall\,\mathbf v_h\in\mathbf{V}_h.
\end{aligned}
\end{equation}
$F_h^{n}(\cdot)$ denotes the discrete fluctuating force at time step $t_n$, taken to be a
mean-zero Gaussian functional on $\mathbf V_h$ with covariance
\begin{equation}
\label{eq:disc-FDT}
\langle F_h^{n}\left(\mathbf v_h\right)\,F_h^{m}(\mathbf w_h)\rangle
=\dfrac{ 2\,k_B T}{\Delta t}\,\; a_h(\mathbf v_h,\mathbf w_h)\,\delta_{nm},
\qquad \forall\,\mathbf v_h,\mathbf w_h\in\mathbf V_h .
\end{equation}

\section{Discrete fluctuation-dissipation balance of the linearized equation}\label{sec:linearization}
\subsection{Linearized equations}\label{sec:lin-scheme}
To highlight the fluctuation--dissipation structure and obtain a linear stochastic system
suitable for covariance analysis, we
linearize \eqref{eq_NS1}--\eqref{eq_NS2} about a homogeneous equilibrium
state. We assume a constant reference density $\rho_0>0$ and vanishing
mean velocity, and write
$$
\rho(\mathbf x,t)=\rho_0+\rho'(\mathbf x,t),\qquad
\mathbf u(\mathbf x,t)=\mathbf u'(\mathbf x,t),\qquad
p(\mathbf x,t)=p_0+p'(\mathbf x,t),
$$
where $(\rho',\mathbf u',p')$ denote small perturbations. We close the
linearized model by assuming an isothermal equation of state and
linearizing at $\rho_0$,
\begin{equation}\label{eq:lin-eos}
p'=\left.\dfrac{\partial p}{\partial \rho}\right|_{\rho=\rho_0}\rho'
=:c_T^2\,\rho',
\end{equation}
with $c_T^2 = R T>0$ the isothermal sound speed. Neglecting higher-order terms
in $(\rho',\mathbf u')$, in particular
$\nabla\!\cdot(\rho'\mathbf u')$ and $\nabla\!\cdot(\rho_0\,\mathbf u'\otimes\mathbf u')$, and using that
$\nabla p_0=0$, we obtain the linearized fluctuating hydrodynamics
system: find $(\rho',\mathbf u')$ such that
\begin{align}
\partial_t\rho' + \rho_0\,\nabla\!\cdot\mathbf u' &= 0,
\label{eq:lin-system1}\\
\rho_0\,\partial_t \mathbf u' + \nabla(c_T^2\rho')
&=
\nabla\!\cdot\!\Big[
 \mu\big(\nabla\mathbf u'+(\nabla\mathbf u')^{\!\top}\big)
 - \dfrac{2}{d}\,\big(\nabla\!\cdot(\mu\,\mathbf u')\big)\mathbf I
\Big]
+ \nabla\!\cdot\widetilde{\boldsymbol{\tau}}.
\label{eq:lin-system2}
\end{align}
For notational simplicity, in the remainder of this section we drop the
primes and interpret $(\rho,\mathbf u)$ as perturbation variables.

As in Section~\ref{weak_formulation}, we write \eqref{eq:lin-system1}--\eqref{eq:lin-system2}
in weak form by testing with functions in $W$ and $V$ and integrating by parts in space. We
again assume boundary conditions such that the boundary contributions vanish. Find
$(\rho(t),\mathbf u(t))\in W\times V$ such that, for almost every $t\in(0,T_f)$,
\begin{align}
(\partial_t\rho,\psi) - \rho_0\,(\mathbf u,\nabla\psi) &= 0,
\qquad \forall\,\psi\in W,
\label{eq:lin-weak-cont}\\
\rho_0\,(\partial_t\mathbf u,\mathbf v)
+ c_T^2\,(\rho,\nabla\!\cdot\mathbf v)
+ a(\mathbf u,\mathbf v)
- F(\mathbf v,t) &= 0,
\qquad \forall\,\mathbf v\in V,
\label{eq:lin-weak-mom}
\end{align}
where $a(\cdot,\cdot)$ is the viscous bilinear form \eqref{eq:viscous-bilinear-form}, and
$F(\cdot,t)$ is defined by \eqref{eq:def-F} with covariance \eqref{eq:weak-covariance}. 

We now discretise \eqref{eq:lin-weak-cont}--\eqref{eq:lin-weak-mom} in time using a
semi-implicit $\theta$--scheme. Given $(\rho^n,\mathbf u^n)\in W_h\times \mathbf V_h$,
find $(\rho^{n+1},\mathbf u^{n+1})\in W_h\times \mathbf V_h$ such that
\begin{align}
\bigg(\dfrac{\rho^{n+1}-\rho^{n}}{\Delta t}, \psi_h\bigg)_{0,h}
- \rho_0\,(\mathbf u^{n,\theta}, \nabla \psi_h)_{0,h}
&= 0,
\,  \forall\,\psi_h\in W_h,
\label{eq:lin-theta-cont}\\[0.4em]
\rho_0\bigg(\dfrac{\mathbf u^{n+1}-\mathbf u^{n}}{\Delta t}, \mathbf v_h\bigg)_{0,h}
+ c_T^2\,(\rho^{n,\theta}, \nabla\!\cdot \mathbf v_h)_{0,h}
&\notag\\
+ a_h(\mathbf u^{n,\theta},\mathbf v_h)
- F_h^{n}(\mathbf v_h)
&= 0,
\,  \forall\,\mathbf v_h\in\mathbf V_h.
\label{eq:lin-theta-mom}
\end{align}
Here $F_h^{n}(\cdot)$ denotes the discrete fluctuating force over the time step
$[t_n,t_{n+1}]$, taken independent across time steps and satisfying the covariance
condition \eqref{eq:disc-FDT}. 
\subsection{Matrix representation}\label{sec:lin-matrix}
We rewrite \eqref{eq:lin-theta-cont}--\eqref{eq:lin-theta-mom} in matrix form by expanding
the discrete unknowns in finite element bases. Let $\{\varphi_i\}_{i=1}^{N_\rho}$ be a basis of
$W_h$ and $\{\boldsymbol{\phi}_j\}_{j=1}^{N_u}$ a basis of $\mathbf V_h$, and write
$$
\rho^n = \sum_{i=1}^{N_\rho}\rho_i^n\,\varphi_i,\qquad
\mathbf u^n = \sum_{j=1}^{N_u}U_j^n\,\boldsymbol{\phi}_j .
$$

Denote by $\boldsymbol{\rho}^n\in\mathbb R^{N_\rho}$ and $\mathbf U^n\in\mathbb R^{N_u}$
the corresponding coefficient vectors. We introduce the mass matrices
\[
(M_\rho)_{ik}:=(\varphi_k,\varphi_i)_{0,h},\qquad
(M_u)_{jl}:=(\boldsymbol{\phi}_l,\boldsymbol{\phi}_j)_{0,h},
\]
the discrete divergence coupling
\[
B_{ij}:=(\nabla\!\cdot \boldsymbol{\phi}_j,\varphi_i)_{0,h},
\]
and the viscous stiffness matrix $K$, defined by its entries $K_{jl} = a_h(\boldsymbol{\phi}_l, \boldsymbol{\phi}_j)$ for all $j, l = 1, \ldots, N_u$. Let $\mathbf y^n :=(\boldsymbol{\rho}^n,\mathbf U^n)^\top$ denote the coefficient vector of
$(\rho^n,\mathbf u^n)\in W_h\times\mathbf V_h$.
Let $\mathbf f^{\,n}\in\mathbb R^{N_u}$ be defined by $(\mathbf f^{\,n})_j := F_h^{n}(\boldsymbol{\phi}_j)$,
$j=1,\ldots,N_u$. Using $M_\rho$, $M_u$, $B$, and $K$, define
\begin{equation}\label{eq:lin-block-matrices}
\mathbf M :=
\begin{pmatrix}
M_\rho & 0\\
0 & \rho_0 M_u
\end{pmatrix},
\qquad
\mathbf D :=
\begin{pmatrix}
0 & \rho_0 B\\
-\,c_T^2 B^{\top} & K
\end{pmatrix},
\qquad
\mathbf g^{\,n}:=
\begin{pmatrix}
0\\
\mathbf f^{\,n}
\end{pmatrix}.
\end{equation}
Here $M_\rho\in\mathbb R^{N_\rho\times N_\rho}$, $M_u,K\in\mathbb R^{N_u\times N_u}$, and $B\in\mathbb R^{N_\rho\times N_u}$. Then the fully discrete $\theta$--scheme can be written as
\begin{equation}\label{eq:lin-theta-block-system}
\mathbf M\,\dfrac{\mathbf y^{n+1}-\mathbf y^{n}}{\Delta t}
+\mathbf D\,\mathbf y^{n,\theta}
=
\mathbf g^{\,n},
\qquad
\mathbf y^{n,\theta}:=\theta\mathbf y^{n+1}+(1-\theta)\mathbf y^{n}.
\end{equation}
The stochastic part satisfies

\begin{equation}\label{eq:lin-theta-block-cov}
\mathbb E\!\left[\mathbf f^{\,n}(\mathbf f^{\,m})^{\top}\right]
=\dfrac{2\,k_B T }{\Delta t}\,K\,\delta_{nm},
\end{equation}

which is the matrix form of \eqref{eq:disc-FDT}.
\subsection{Discrete fluctuation-dissipation balance}
Starting from \eqref{eq:lin-theta-block-system}, the $\theta$--scheme can be expressed equivalently as

\begin{equation}\label{eq:theta_one_step}
\left(\mathbf M+\theta \Delta t\,\mathbf D\right)\mathbf y^{n+1}
=
\left\{\mathbf M-(1-\theta)\Delta t\,\mathbf D\right\}\mathbf y^{n}
+ \Delta t \mathbf g^{\,n}.
\end{equation} 

Assume that $\mathbf y^{n}$ admits a stationary covariance
$$
\mathbf C := \mathrm{Cov}(\mathbf y^{n}) = \mathrm{Cov}(\mathbf y^{n+1}),
$$
and that the forcing is white in time and independent of the current state (additive), 
\begin{equation}\label{eq:white_noise_assumptions}
\mathbb E\!\left[\mathbf f^{\,n}(\mathbf f^{\,m})^{\top}\right]
=\mathbf C_{ff}\,\delta_{nm},
\qquad
\mathrm{Cov}(\mathbf y^{n},\mathbf f^{\,n})=\mathbf 0.
\end{equation}
Equivalently,
\[
\mathbb E\!\left[\mathbf g^{\,n}(\mathbf g^{\,m})^{\top}\right]
=\mathbf C_{gg}\,\delta_{nm},
\qquad
\mathbf C_{gg}:=
\begin{pmatrix}
0 & 0\\
0 & \mathbf C_{ff}
\end{pmatrix}.
\]

Taking the covariance of both sides of \eqref{eq:theta_one_step} yields, after expansion and simplification,
\begin{equation}\label{eq:disc_fdt_image_style}
\mathbf D\,\mathbf C\,\mathbf M^{\top}
+\mathbf M\,\mathbf C\,\mathbf D^{\top}
+\Delta t\,(2\theta-1)\,\mathbf D\,\mathbf  C\,\mathbf D^{\top}
=
\mathbf C_{gg}.
\end{equation}
For the Euler choices $\theta=0$ (explicit) and $\theta=1$ (implicit), the
$\Delta t$--dependent contribution $\Delta t(1-2\theta)\,\mathbf D\,\mathbf C\,\mathbf D^{\top}$
does not vanish in \eqref{eq:disc_fdt_image_style}. As a result,
matching the discrete fluctuation--dissipation balance to a prescribed stationary
covariance typically requires a stability-type condition that couples the time step
$\Delta t$ to the spatial discretisation, as discussed
in \cite{MartinezLeraDeCorato2024}. Choosing $\theta=\tfrac12$ (Crank--Nicolson) makes the third term in
\eqref{eq:disc_fdt_image_style} vanish, and the discrete balance reduces to
\begin{equation}\label{eq:discFDT_CN}
 \mathbf D\,\mathbf C\,\mathbf M^{\top}+\mathbf M\,\mathbf C\,\mathbf D^{\top}
= \mathbf C_{gg}.   
\end{equation}
This is the stationary covariance balance associated with the fully discrete linear system \eqref{eq:lin-theta-block-system}.

We now verify the Crank--Nicolson discrete fluctuation--dissipation relation. Using the block matrices
$\mathbf M$ and $\mathbf D$ from \eqref{eq:lin-block-matrices}, consider the block-diagonal covariance
\begin{equation}\label{eq:C-choice}
\mathbf C
= k_B T
\begin{pmatrix}
\displaystyle \frac{\rho_0}{c_T^2}\,M_\rho^{-1} & 0\\[0.4em]
0 & \displaystyle \frac{1}{\rho_0}\,M_u^{-1}
\end{pmatrix}.
\end{equation}
Substituting \eqref{eq:lin-block-matrices} and \eqref{eq:C-choice} into \eqref{eq:discFDT_CN} shows that the conservative coupling cancels, so \eqref{eq:C-choice} is the stationary covariance for Crank--Nicolson.

In the continuum isothermal setting, equilibrium fluctuations are spatially local,
so the covariance operator is diagonal in physical space. In contrast, for standard
continuous finite elements the $L^2$ mass matrices $M_\rho$ and $M_u$ are typically
non-diagonal, reflecting the overlap of basis functions. Consequently, the covariance \eqref{eq:C-choice} inherits off-diagonal entries through
$M_\rho^{-1}$ and $M_u^{-1}$. These mesh-scale
correlations are not physical; they arise from the representation of the fields in a nodal basis together with the consistent $L^2$ inner product, and should be understood as a structural consequence of the discretisation rather than a convergence error.

To mitigate this effect, we use a nodal quadrature rule to define the discrete inner
product, leading to a diagonal lumped mass matrix. This removes the cross-covariances in the discrete equilibrium covariance while introducing only a mild loss of quadrature accuracy, which is negligible in the examples considered in this work. A simple
one-dimensional illustration of how nodal quadrature eliminates the off-diagonal entries of the mass matrix is given in \ref{Appendix B}. The use of mass lumping effectively makes the finite element discretisation behave similarly to a finite volume method in terms of fluctuation statistics, while maintaining the advantages of a variational formulation.
Related strategies for stochastic PDEs based on modifying the finite element space or basis have also been considered in the literature \citep{MartinezLeraDeCorato2024,Djurdjevac2026}.


\section{Numerical results}\label{sec:numerical_results}

This section presents numerical validation of the fully discrete scheme developed in Sections~\ref{Variational_formulation} and \ref{sec:linearization}. Our primary objective is to confirm that the method correctly reproduces the equilibrium fluctuation statistics prescribed by the fluctuation-dissipation theorem. We achieve this by simulating an isothermal fluid in a near-equilibrium state and comparing the empirical variances of density and velocity fields against their theoretical from statistical mechanics \citep{de2006hydrodynamic}.

 We test the method across multiple mesh resolutions and different element types, spatial dimensions, and time-step sizes, using long final times $T_f$, to verify that the discrete stochastic forcing and viscous dissipation satisfy the correct fluctuation--dissipation balance.

Let $|\mathcal T_h^i|$ denote the $d$--dimensional measure of a mesh element: length in $d=1$,
area in $d=2$, and volume in $d=3$. For elementwise  fluctuations, the equilibrium variance of the density satisfies
\begin{equation}\label{eq:var_rho_theory}
\langle \delta \rho^2 \rangle = \dfrac{k_B T \langle \rho \rangle}{c_T^2\,|\mathcal T_h^i|},
\end{equation}
and, for the velocity,
\begin{equation}\label{eq:var_u_theory}
\langle \delta u^2 \rangle = \dfrac{k_BT}{\langle \rho \rangle\,|\mathcal T_h^i|}.
\end{equation}
Equations \eqref{eq:var_rho_theory}--\eqref{eq:var_u_theory} are the local form of the
discrete equilibrium covariance \eqref{eq:C-choice} with $\rho_0=\langle\rho\rangle$ is the equilibrium density,
and $M_\rho$, $M_u$ are the mass matrices associated with $(\cdot,\cdot)_{0,h}$;
the elementwise variance levels follow from the diagonal scaling of $M^{-1}$,
namely $(M^{-1})_{ii}\sim |\mathcal T_h^i|^{-1}$. 

In the results below, we use boundary conditions for which the boundary terms in the weak
formulation vanish: periodic boundaries in 1D and 2D, and in 3D periodicity on one pair of
opposite faces with homogeneous Neumann conditions on the remaining
faces. The mesh size $h$ is chosen so that each mesh element contains a sufficient number of molecules, in accordance with the requirements of hydrodynamic coarse-graining, see \cite{RussoPerezDuranYatsyshinCarrilloKalliadasis}. 
Unless otherwise specified, we adopt the non-dimensional parameters ($\mu = 10$), ($T = 1.2$), and ($c_T^2 = 1.2$), expressed in Lennard--Jones reduced units \cite{gallo2018thermally}.

We perform all simulations using FEniCSx \citep{Baratta2023DOLFINx,Scroggs2022DofMaps,Scroggs2022BasixJOSS,Alnaes2014UFL} on a laptop equipped with Intel\textsuperscript{\tiny\textregistered} Core\texttrademark\ i9-13950HX. At each time step, the linear systems arising from the variational formulation are all assembled and solved with PETSc, \citep{Dalcin2011ParallelPython,Balay2025PETScManual}. 

\subsection{One-dimensional example}

\begin{figure}[htbp]
\centering
\begin{minipage}{0.48\textwidth}
    \centering
    \includegraphics[width=\linewidth]{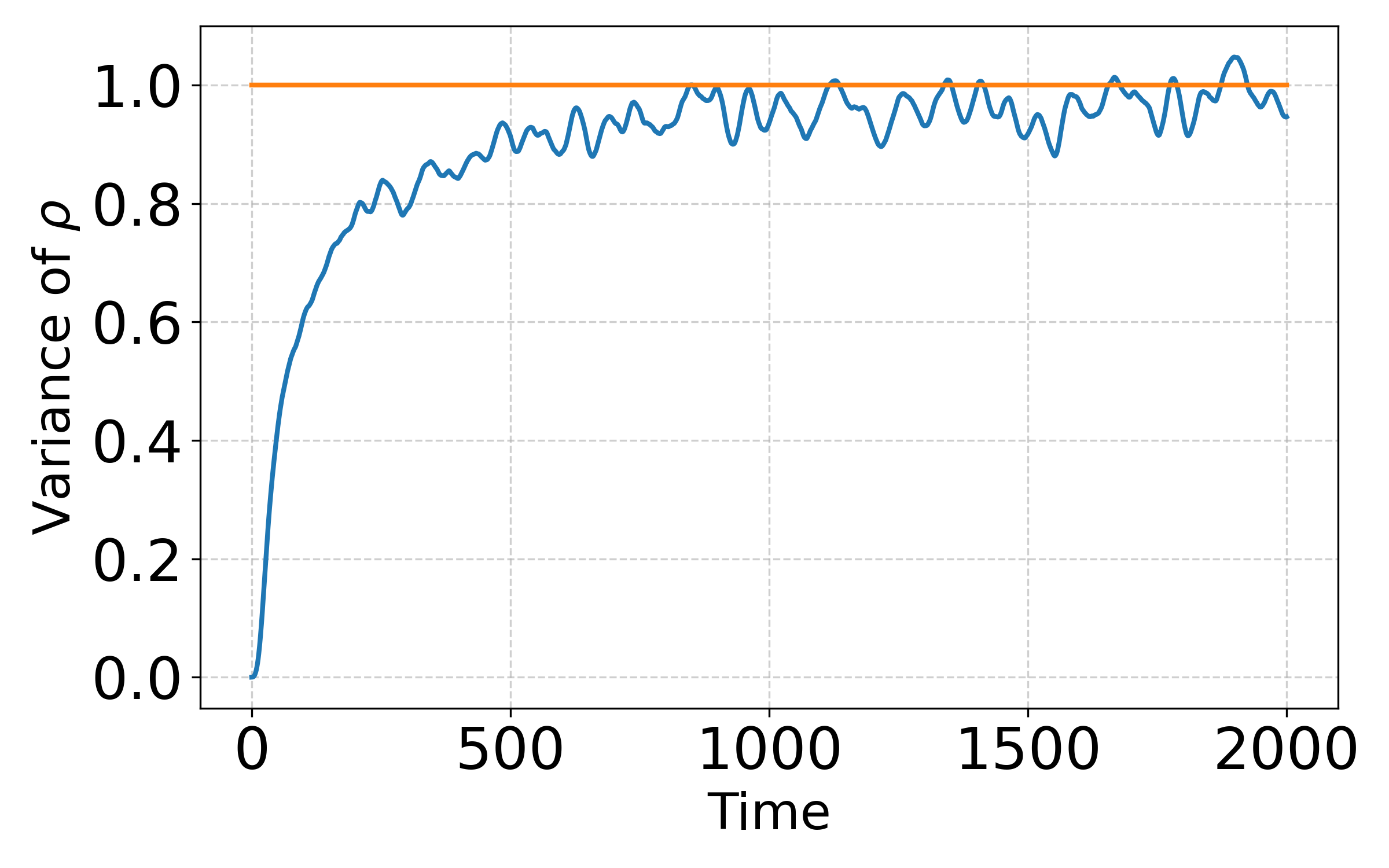}
    
\end{minipage}\hfill
\begin{minipage}{0.48\textwidth}
    \centering
    \includegraphics[width=\linewidth]{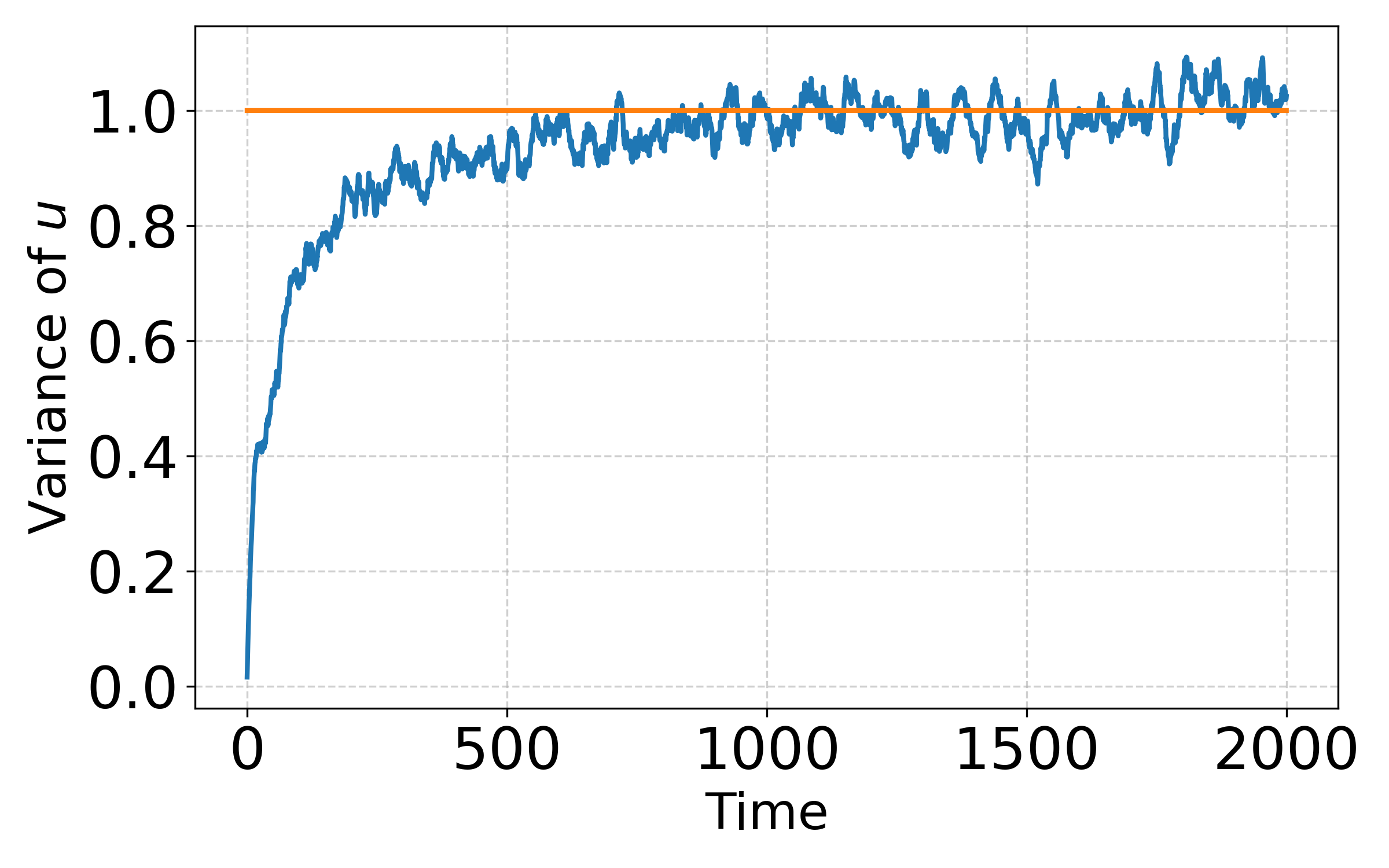}
    
\end{minipage}
\par\smallskip
{\footnotesize\centering $\Delta t = 0.4,\quad h=20$.\par}
\vspace{0.8em}

\begin{minipage}{0.48\textwidth}
    \centering
    \includegraphics[width=\linewidth]{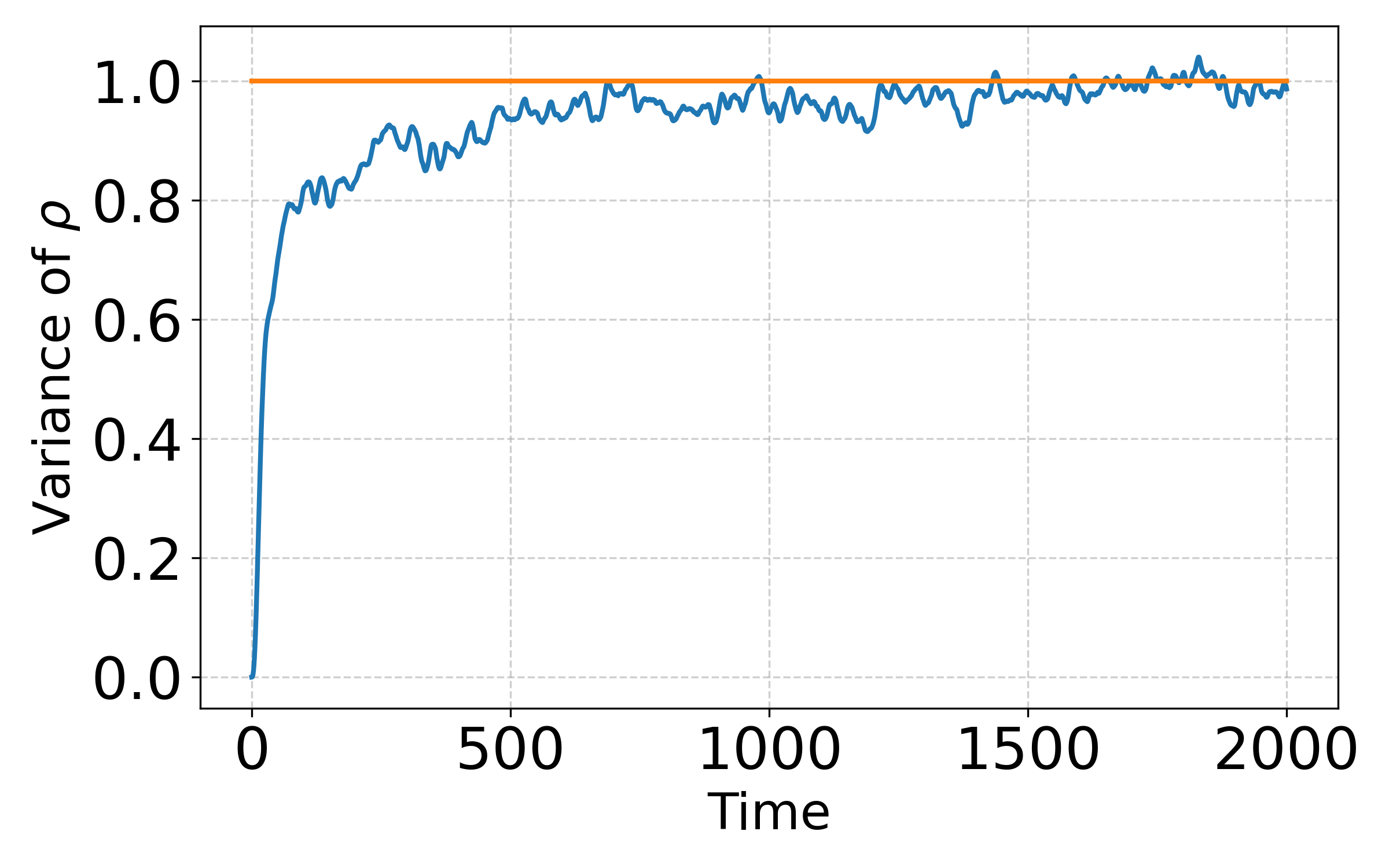}
   \end{minipage}\hfill
\begin{minipage}{0.48\textwidth}
    \centering
    \includegraphics[width=\linewidth]{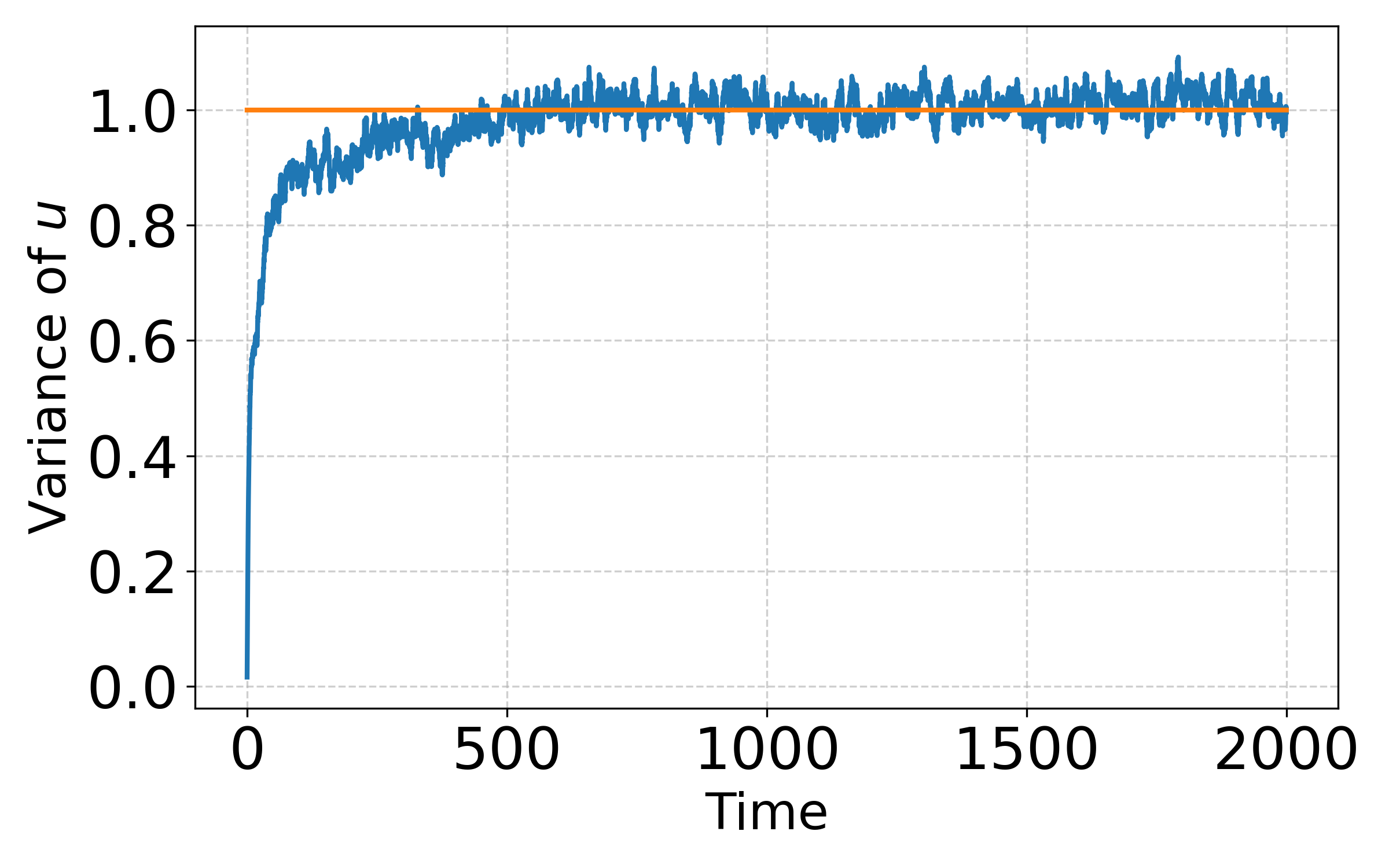}
    \end{minipage}
{\footnotesize\centering $\Delta t =0.1,\quad h=10$.\par}
\vspace{0.8em}

\begin{minipage}{0.48\textwidth}
    \centering
    \includegraphics[width=\linewidth]{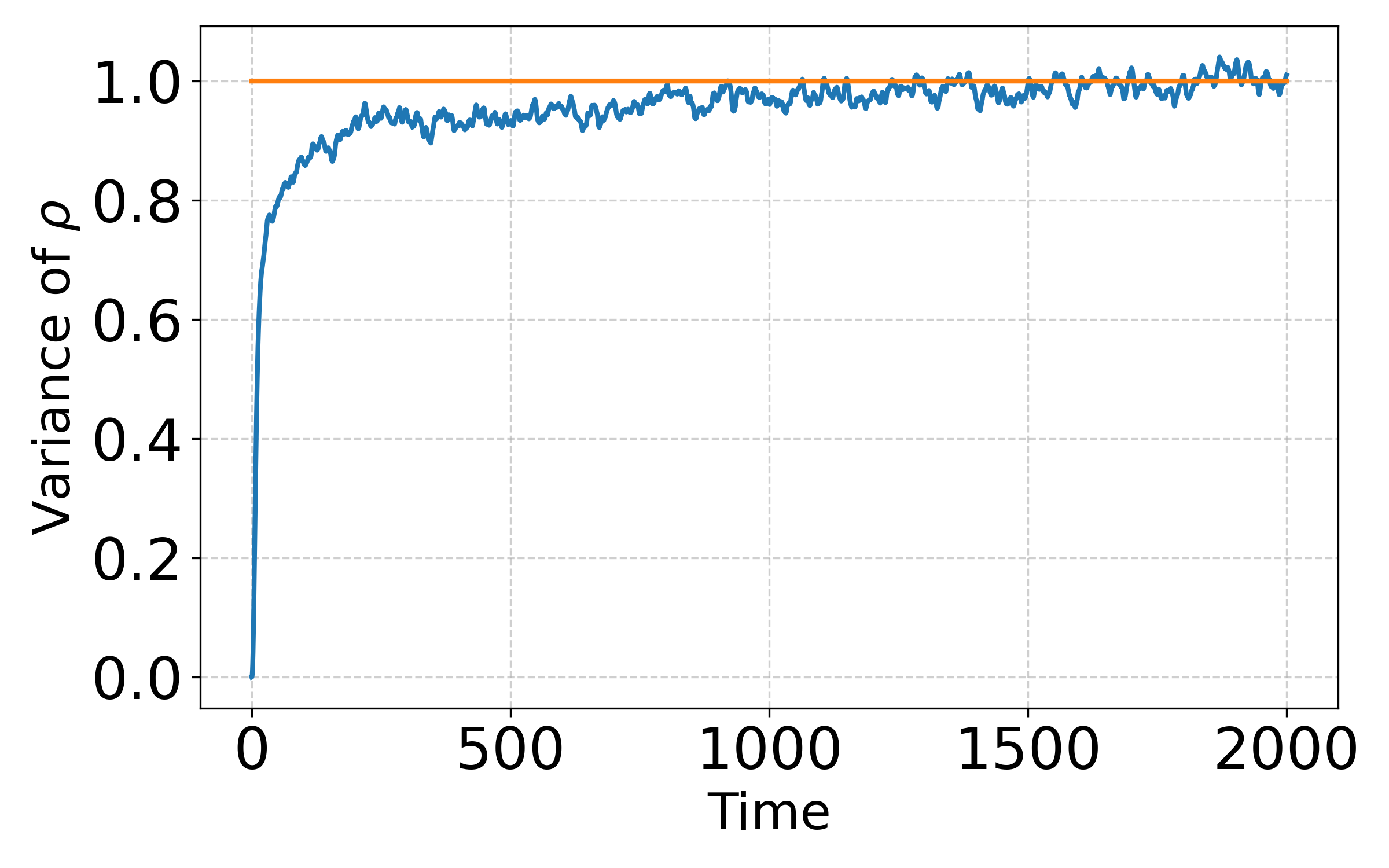}
   \end{minipage}\hfill
\begin{minipage}{0.48\textwidth}
    \centering
    \includegraphics[width=\linewidth]{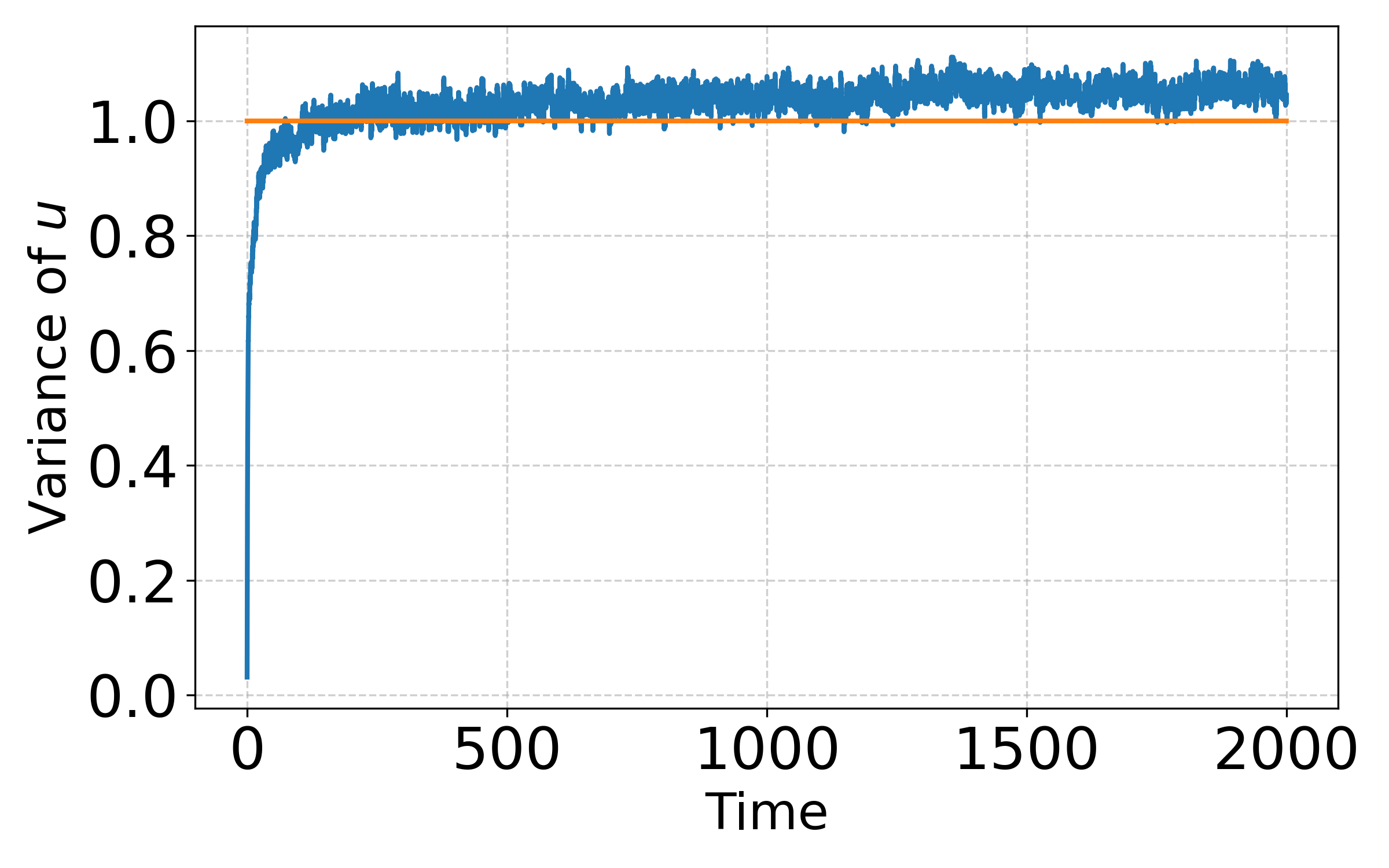}
    \end{minipage}
{\footnotesize\centering $\Delta t =0.05, \quad h=5$.\par}

\caption{One-dimensional example: time evolution of the normalised sample variances of the density $\rho$ (left) and velocity $u$ (right) for $\Delta t\in\{0.1,\,0.05,\,0.025\}$, and $h \in\{20,10,5\}$ (top to bottom). The horizontal orange line indicates the equilibrium reference level $1$.}
\label{fig:variance_dt}
\end{figure}

We consider a stable liquid with $\langle \rho \rangle = 5$, for several time steps $\Delta t\in\{0.05,0.1,0.4\}$ and mesh sizes $h \in\{5,10,15,20\}$. Figure~\ref{fig:variance_dt} shows the normalised sample variances of the density and the velocity, i.e., $\hat{s}^2_\rho/\langle\delta\rho^2\rangle$,$\hat{s}^2_u/\langle\delta u^2\rangle$. 
where the sample variance $\hat{s}^2_X$ computed from $N$ samples is
\begin{equation}\label{eq:sample-variance}
\hat{s}^2_X=\dfrac{1}{N-1}\sum_{n=1}^N\big(X_{n}-\bar X\big)^2,
\qquad
\bar X=\dfrac{1}{N}\sum_{n=1}^N X_n.
\end{equation}
Here $X$ denotes either $\rho$ or $u$, and $\bar X$ is the corresponding sample mean.

The sample variances of both the density and velocity evolve from the initial condition, when the initial fields contain no thermal fluctuation content and then increase to approach the theoretical equilibrium level. Once equilibrium is reached, the sample variances fluctuate around the reference value  due to finite-time sampling of a stochastic process. The amplitude of these fluctuations decreases from the top row to the bottom row because spatial refinement
increases the number of degrees of freedom over the fixed domain: for $\mathcal{T}_h=[a,b]$,
the number of elements scales like $N \approx (b-a)/h$, so halving $h$ approximately doubles $N$.

The density variance appears noticeably more smooth than the velocity variance in Figure \ref{fig:variance_dt}. This behavior is expected since the underlying stochastic path of the velocity field is directly driven by the Gaussian white noise, whereas the density field is driven by the temporal integration of these velocity fluctuations. This in turn implies that the velocity itself is a Markov process in that the stochastic increments are uncorrelated, but the density alone is non-Markov in that this process has memory which manifests as temporally correlated stochastic increments. At equilibrium, where linearised fluctuating hydrodynamics is valid, both the density and velocity fields are Gaussian with the variances being given by (\ref{eq:var_rho_theory}) and (\ref{eq:var_u_theory}). Thus, if one takes an infinite number of samples to construct the variance the equilibrium variance would look completely flat. However, as this is not the case, we observe the variance plots in Figure \ref{fig:variance_dt}, with the temporally smoother appearance of the density variance coming from the temporal continuity in the  first derivative of the density field.

Figure~\ref{fig:variance_dt} also shows that decreasing $\Delta t$ delays the apparent approach to the equilibrium level (from $t=250$ to $t=1000$ in these runs), an effect that is especially visible in the density variance and less pronounced for the velocity.

Table~\ref{tab:var_refinement} reports the time-averaged variance estimator $\langle \widehat{s}^{\,2}\rangle$ computed by averaging $\widehat{s}^{\,2}(t_n)$ over $t_n\in[T_{\mathrm{eq}},T_f]$.  Consistent with the theoretical scaling $\langle\delta(\cdot)^2\rangle \propto |\mathcal T_h^i|^{-1}$, halving the mesh size approximately doubles the equilibrium variance, which is reflected in the estimated values. The empirical rates $p_u$ and $p_\rho$ are close to one, confirming this factor of two increase per refinement level. The numerical results shown in Figure \ref{fig:variance_dt} and Table~\ref{tab:var_refinement} are obtained using Crank–Nicolson; using explicit Euler time integration yields similar results, which are not reported here.


\begin{table}[htbp]
\centering
\caption{One-dimensional example: average sample variances  $\langle\widehat{s}_u^{\,2}\rangle$ and $\langle\widehat{s}_\rho^{\,2}\rangle$ and the corresponding theoretical values
$\langle \delta u^2\rangle$ and $\langle \delta\rho^2\rangle$. 
Relative errors are $\big(\widehat{s}^{\,2}-\langle\cdot\rangle\big)/\langle\cdot\rangle\times 100\%$,
and the empirical rate is
$
p=\log\!\big(\widehat{s}^{\,2}_{\mathrm{finer}}/\widehat{s}^{\,2}_{\mathrm{coarser}}\big)/
\log\!\big(h_{\mathrm{coarser}}/h_{\mathrm{finer}}\big).
$}
\label{tab:var_refinement}
\setlength{\tabcolsep}{5pt}
\begin{tabular}{
  c
  S[table-format=1.6]
  S[table-format=1.2] 
  S[table-format=+2.2]
  S[table-format=1.6] 
  S[table-format=1.3] 
  S[table-format=+2.2]
  S[table-format=1.3] 
  S[table-format=1.3]
}
\toprule
& \multicolumn{3}{c}{Velocity} & \multicolumn{3}{c}{Density} & \multicolumn{2}{c}{Rate} \\
\cmidrule(lr){2-4}\cmidrule(lr){5-7}\cmidrule(lr){8-9}
{$h$} &
{$\langle\widehat{s}_u^{\,2}\rangle$ } & {$\langle \delta u^2\rangle$} & {Error (\%)} &
{$\langle\widehat{s}_\rho^{\,2}\rangle$} & {$\langle \delta\rho^2\rangle$} & {Error (\%)} &
{$p_\rho$} & {$p_u$} \\
\midrule
20.0  & 0.011493 & 0.012 & -4.23 & 0.231219 & 0.25 & -7.51 & \multicolumn{1}{c}{---} & \multicolumn{1}{c}{---} \\
15.0  & 0.015446 & 0.016 & -3.44 & 0.306861 & 0.33 & -7.92 & 0.985 & 1.028 \\
10.0  & 0.023701 & 0.024 & -1.24 & 0.471614 & 0.50 & -5.68 & 1.059 & 1.055 \\
 5.0  & 0.049642 & 0.048 & +3.42 & 0.959401 & 1.00 & -4.06 & 1.025 & 1.067 \\
\bottomrule
\end{tabular}
\end{table}

\subsection{Multi-dimensional examples}

\begin{figure}[htbp]
  \centering
  \begin{subfigure}[t]{0.48\textwidth}
    \centering
    \includegraphics[scale=0.8,width=\linewidth]{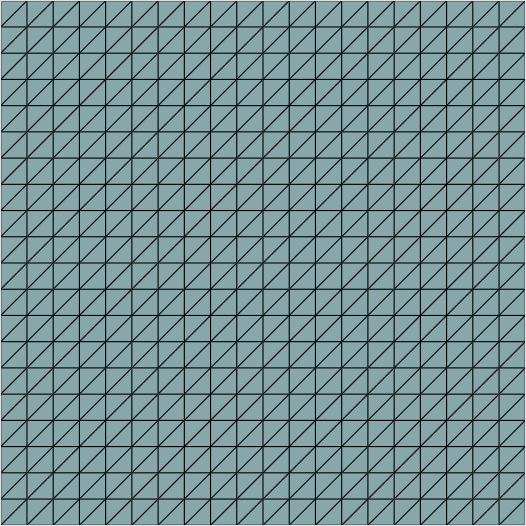}
    \caption{Triangular mesh.}
    \label{fig:mesh-tri}
  \end{subfigure}\hfill
  \begin{subfigure}[t]{0.48\textwidth}
    \centering
    \includegraphics[scale=0.8,width=\linewidth]{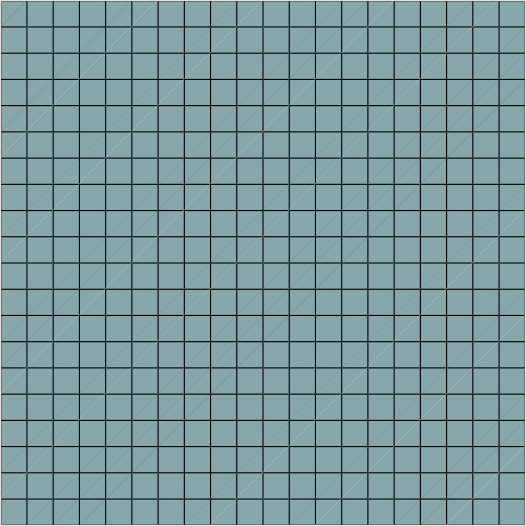}
    \caption{Quadrilateral mesh.}
    \label{fig:mesh-quad}
  \end{subfigure}
  \caption{2D meshes used in the experiments.}
  \label{fig:meshes-2d}
\end{figure}

\begin{figure}[htbp]
\centering
\begin{minipage}{0.48\textwidth}
\centering
\includegraphics[width=\linewidth]{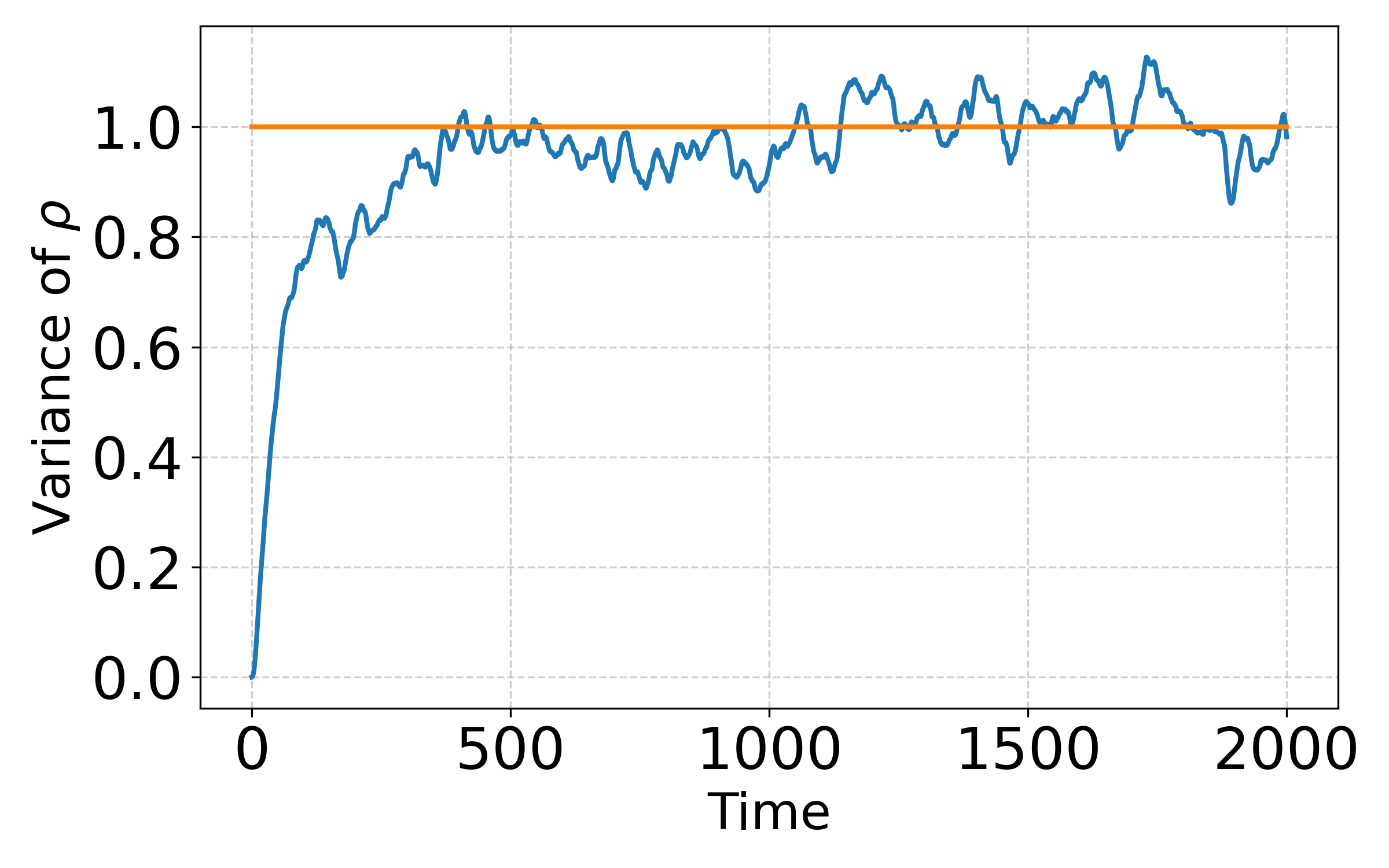}
\end{minipage}\hfill 
\begin{minipage}{0.48\textwidth}
\centering
\includegraphics[width=\linewidth]{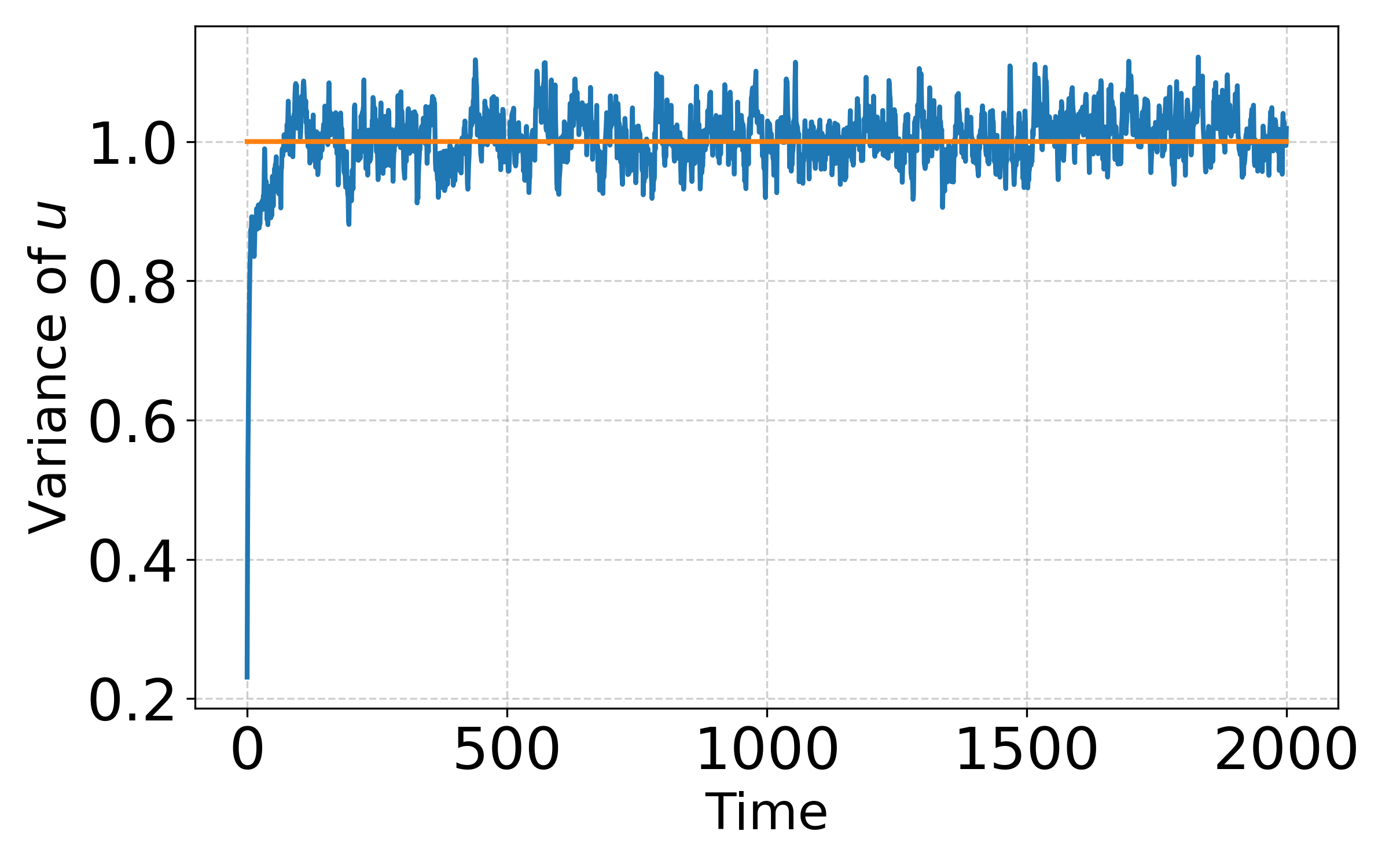}
\end{minipage}

\par\smallskip
{\footnotesize\centering $\Delta t=0.8,\; \mathcal{T}^i_h=374.6$.\par}

\vspace{0.8em}

\begin{minipage}{0.48\textwidth}
\centering
\includegraphics[width=\linewidth]{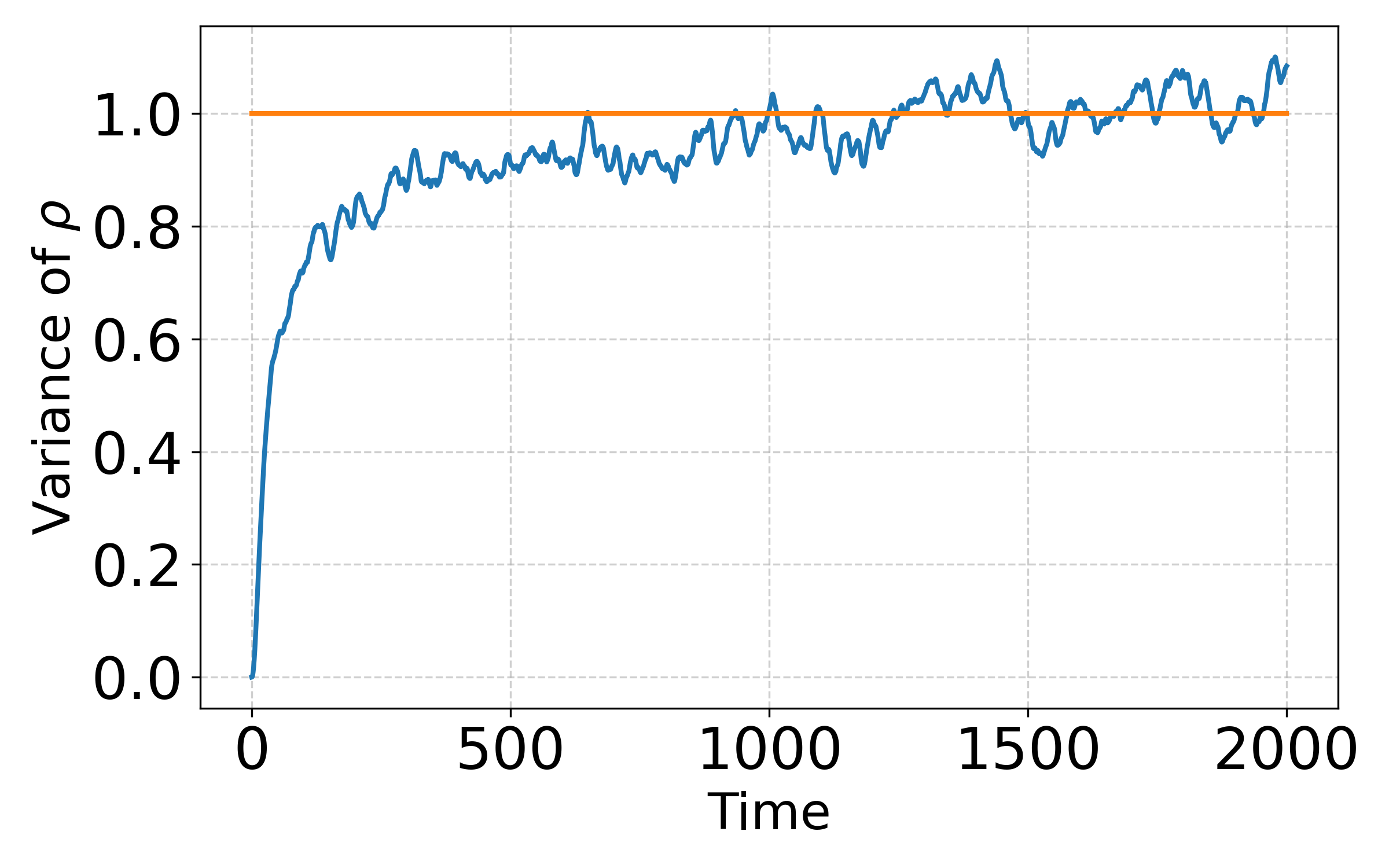}
\end{minipage}\hfill
\begin{minipage}{0.48\textwidth}
\centering
\includegraphics[width=\linewidth]{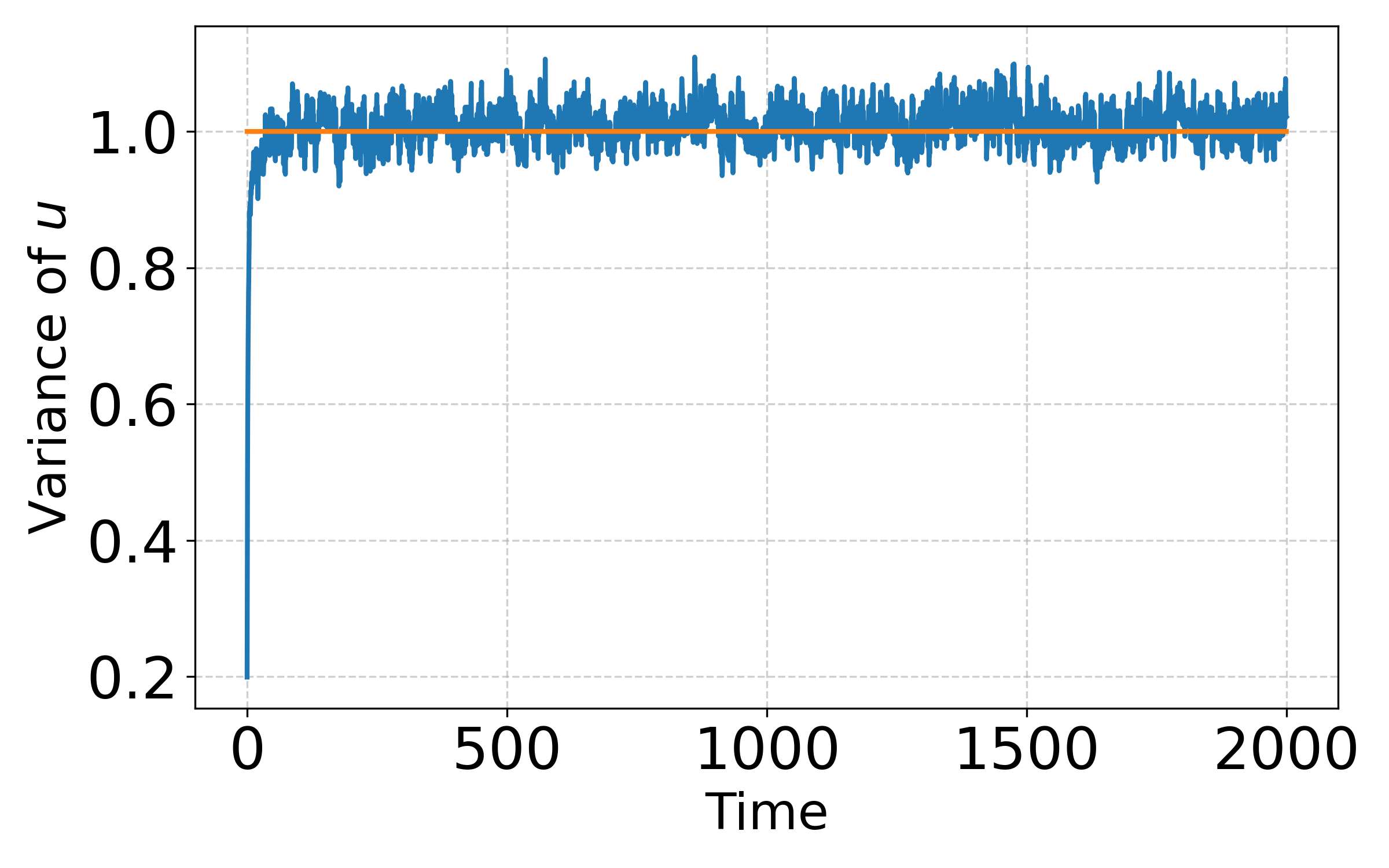}
\end{minipage}

\par\smallskip
{\footnotesize\centering $\Delta t=0.4,\; \mathcal{T}^i_h=214.1$.\par}

\vspace{0.8em}

\begin{minipage}{0.48\textwidth}
\centering
\includegraphics[width=\linewidth]{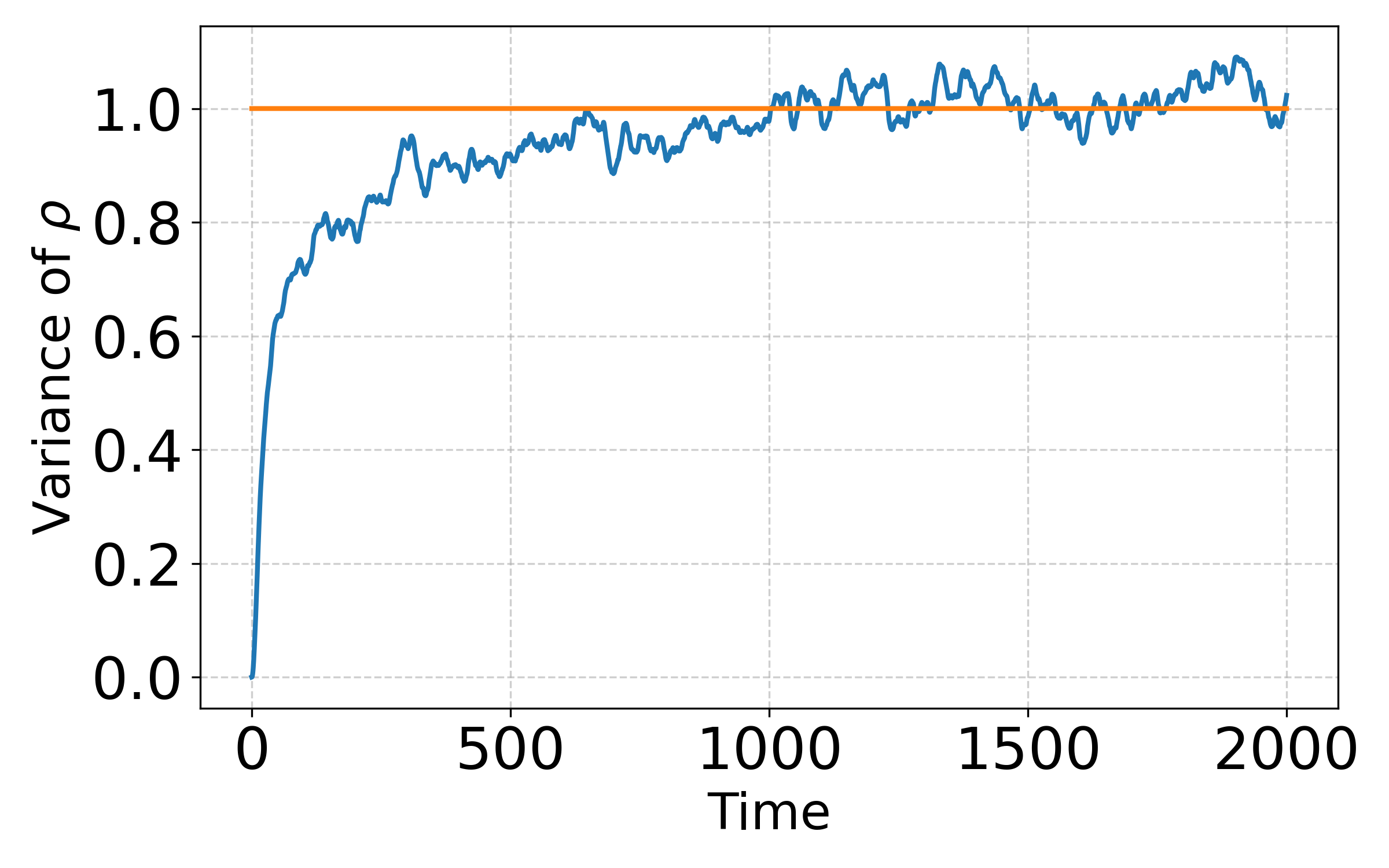}
\end{minipage}\hfill
\begin{minipage}{0.48\textwidth}
\centering
\includegraphics[width=\linewidth]{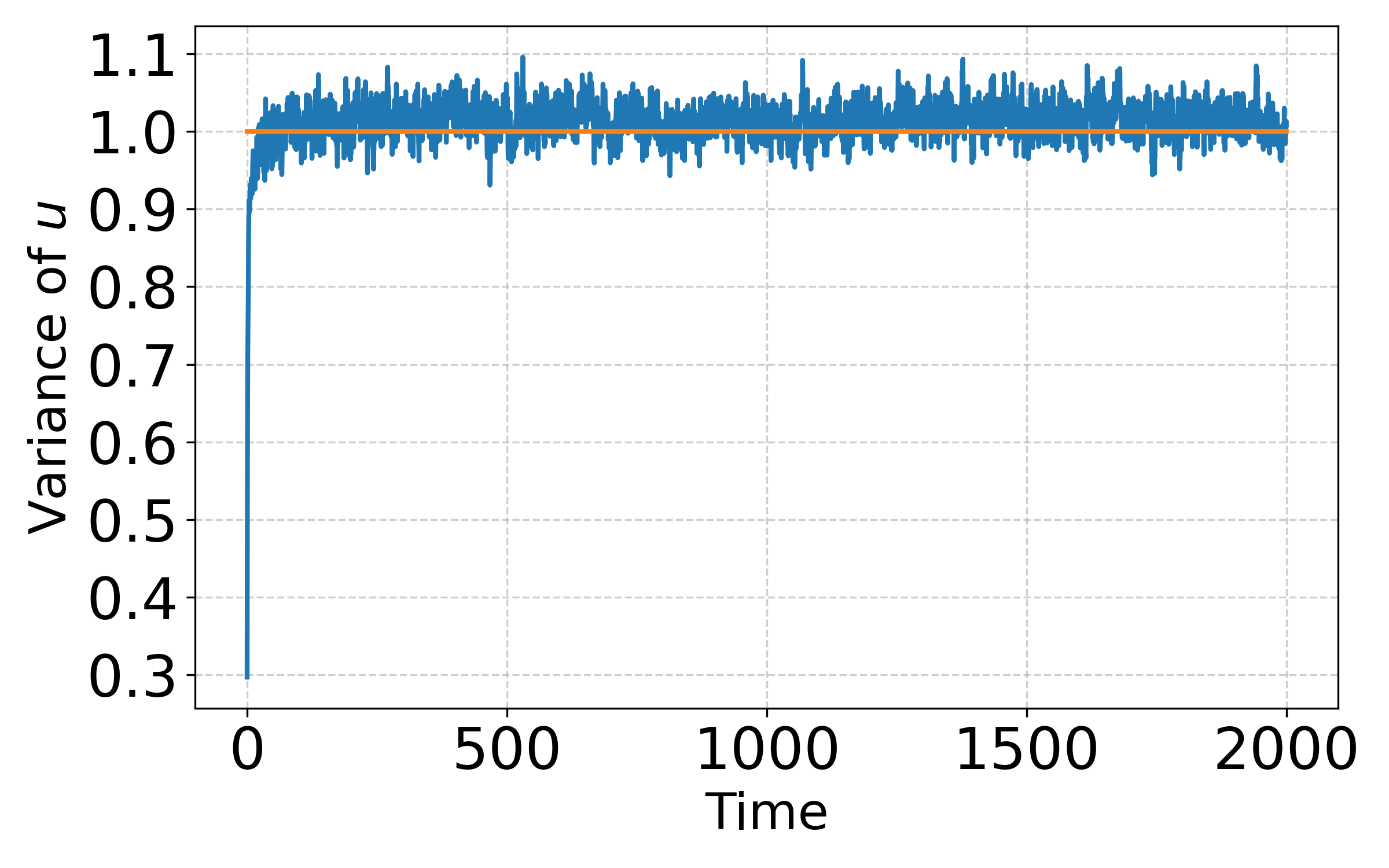}
\end{minipage}

\par\smallskip
{\footnotesize\centering $\Delta t=0.2,\; \mathcal{T}^i_h=138.4$.\par}

\caption{Two-dimensional example: time evolution of the normalised sample variances of the density $\rho$ (left) and velocity $u$ (right) for $\Delta t\in\{0.2,\,0.4,\,0.8\}$ and $\mathcal{T}^i_h\in\{374.6,214.1,138.4\}$ (top to bottom). The horizontal orange line indicates the equilibrium reference level $1$.}
\label{fig:variance_dt_h_2d}
\end{figure}

First, we consider the two-dimensional equilibrium test case and report results for several time steps and mesh sizes on both triangular and quadrilateral grids, see Figure~\ref{fig:meshes-2d}.
Figure~\ref{fig:variance_dt_h_2d} shows the normalised variance estimates for increasing mesh resolution on a triangular mesh.
Most of the observations from the one-dimensional study remain valid in two dimensions. In particular, the variance estimates approach the equilibrium level after a transient whose duration depends on both $\Delta t$ and the mesh size, and, once equilibrium is reached, the trajectories fluctuate around $1$ due to finite-time sampling. Note that in 2D the equilibrium
variances themselves are smaller because \eqref{eq:var_rho_theory}--\eqref{eq:var_u_theory} depend on $|\mathcal{T}_h^i|^{-1}$ and $|\mathcal{T}_h^i|$ is an area, so the variance decreases for coarser mesh size, see Table~\ref{tab:var_dt05_h_2d}. 

\begin{figure}[htbp]
\centering
\begin{minipage}{0.48\textwidth}
\centering
\includegraphics[width=\linewidth]{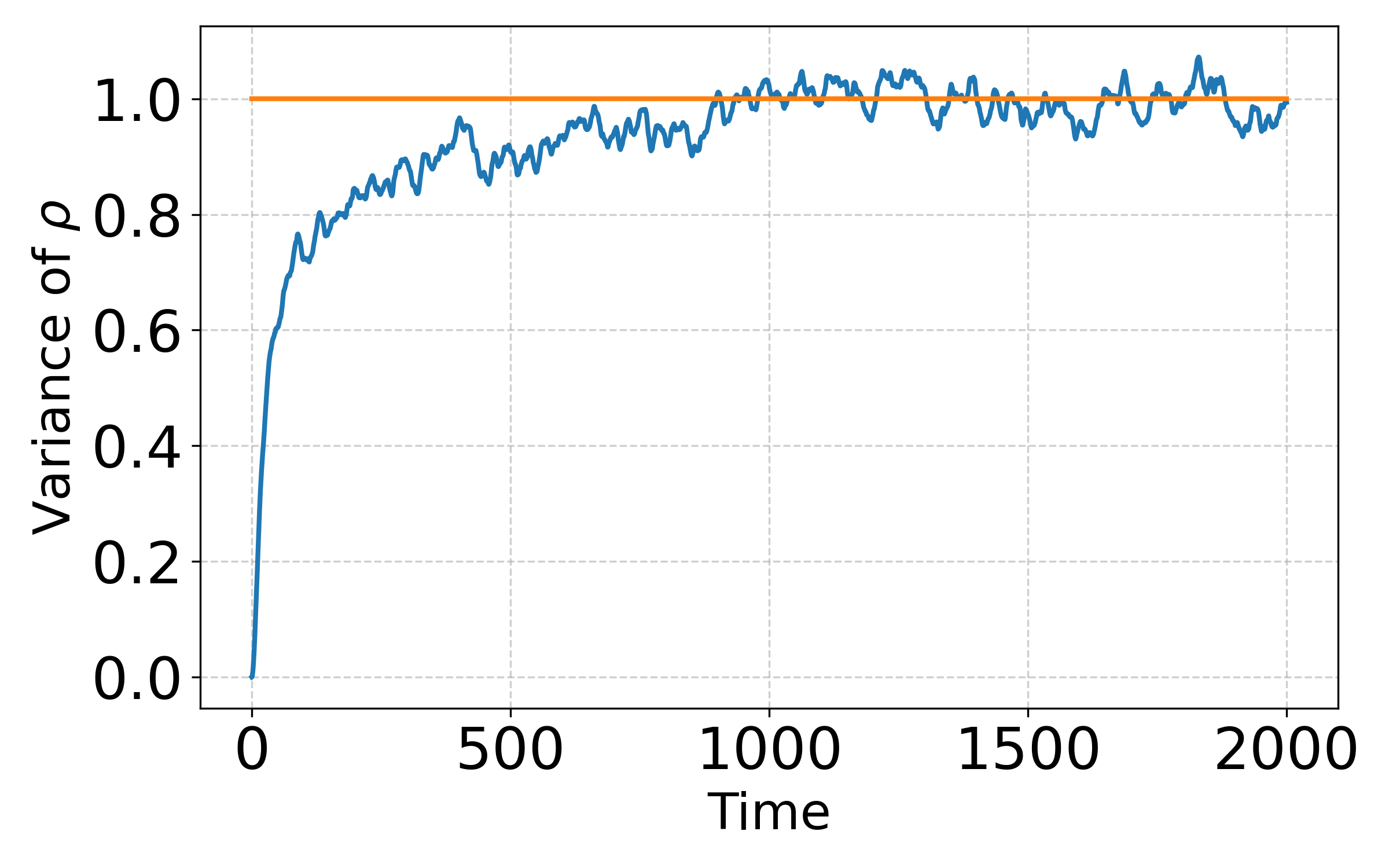}
\end{minipage}\hfill
\begin{minipage}{0.48\textwidth}
\centering
\includegraphics[width=\linewidth]{plots/plots_2d/2d_triangle_var_u_plot_styled_dt=0.4_0.6_138.40830449826993_2000.0.png}
\end{minipage}

\par\smallskip
{\footnotesize\centering $\Delta t=0.4,\; \mathcal{T}^i_h=138.4$.\par}

\vspace{0.8em}

\begin{minipage}{0.48\textwidth}
\centering
\includegraphics[width=\linewidth]{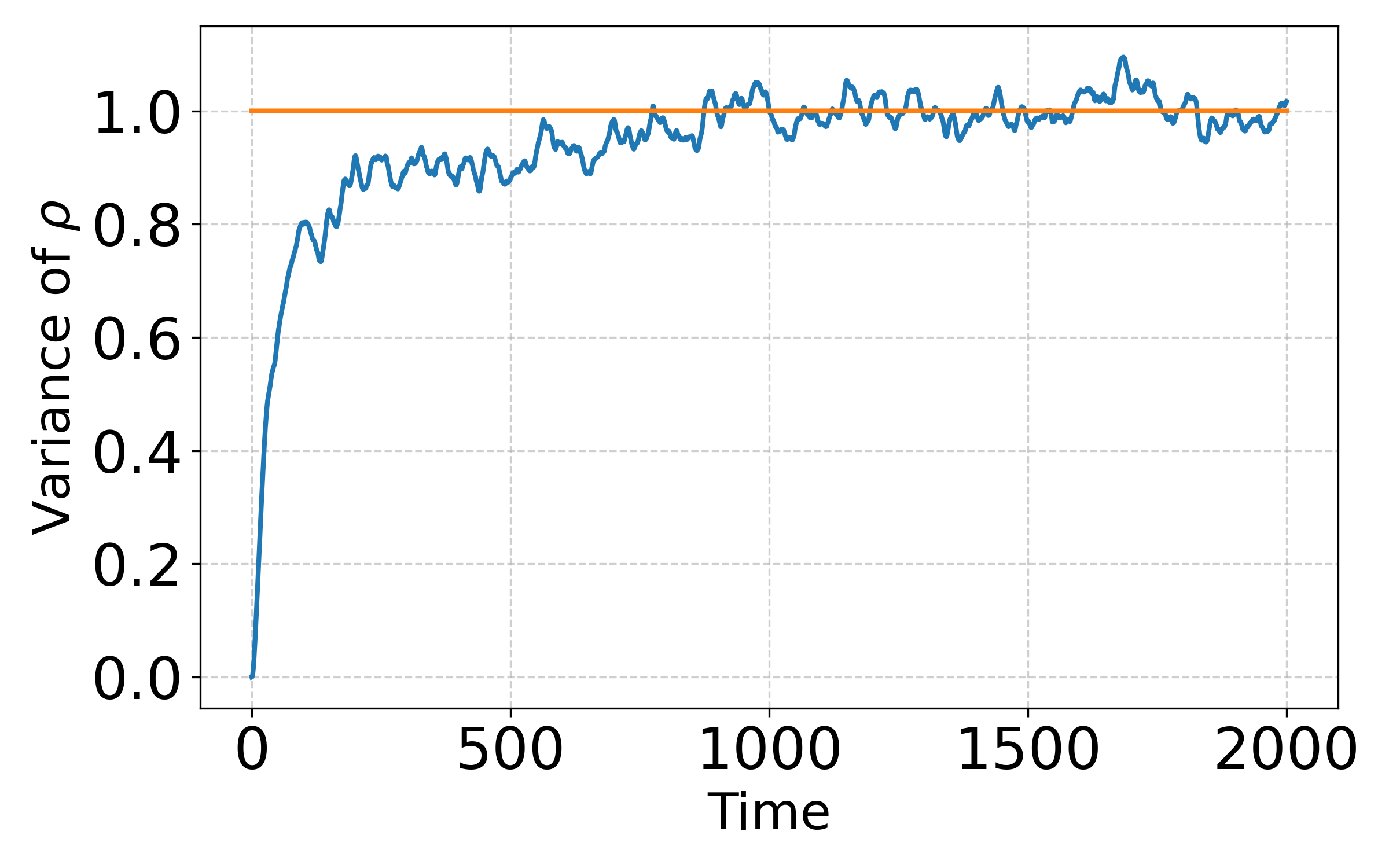}
\end{minipage}\hfill
\begin{minipage}{0.48\textwidth}
\centering
\includegraphics[width=\linewidth]{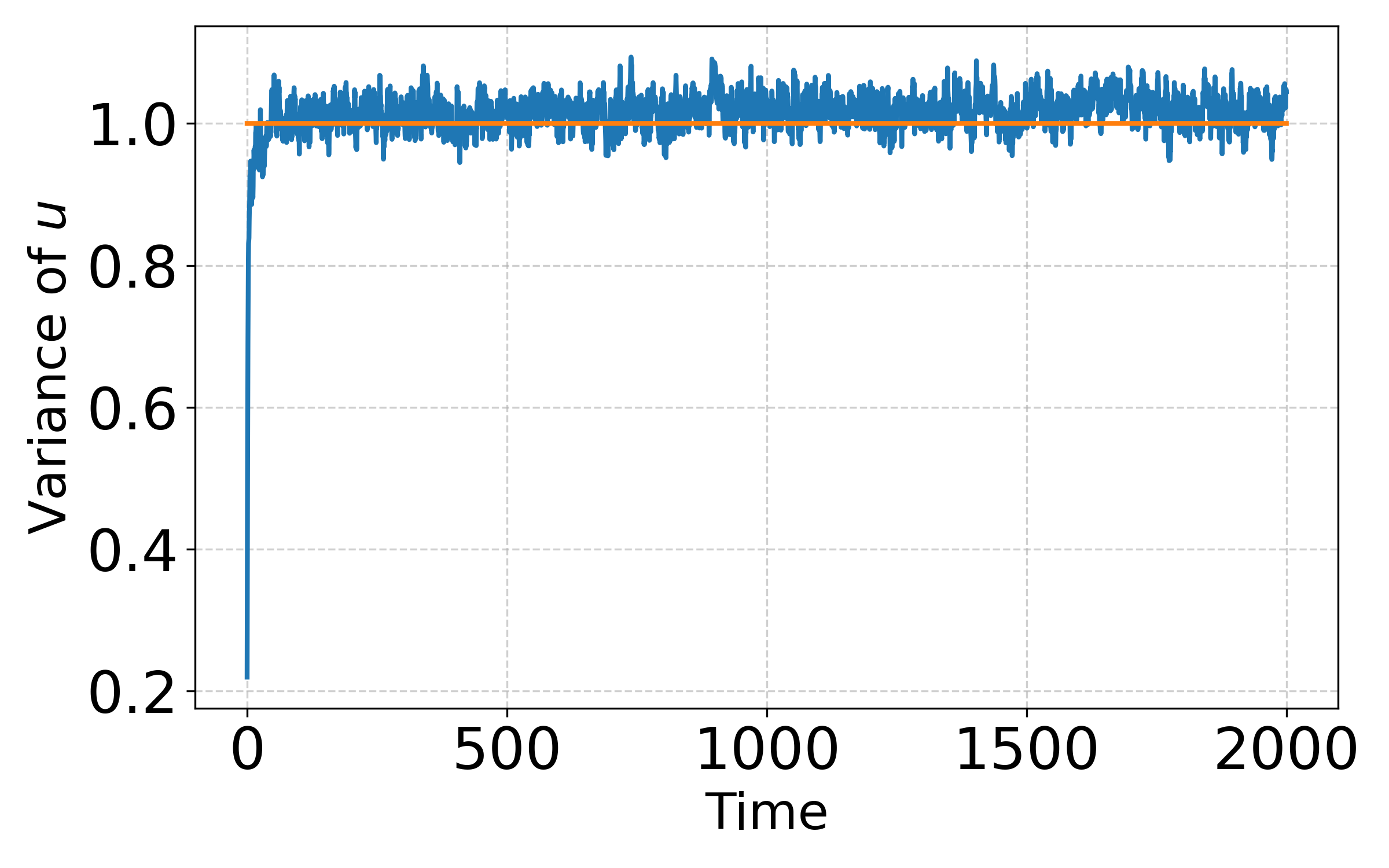}
\end{minipage}

\par\smallskip
{\footnotesize\centering $\Delta t=0.4,\; \mathcal{T}^i_h=138.4$.\par}

\caption{Two-dimensional example: triangular mesh (top row); quadrilateral mesh (bottom row); time evolution of the normalised sample variances of the density $\rho=0.6$ (left) and velocity $u$ (right) for $\Delta t=0.4$ and $h=138.6$. The horizontal orange line indicates the equilibrium reference level $1$.}
\label{fig:mesh_var_2d}
\end{figure}

Comparing triangular and quadrilateral meshes at the same $(h,\Delta t)$, we observe no major difference in the normalised variance estimates; see Figure~\ref{fig:mesh_var_2d}. This comparison suggests that the element shape has no major impact on the estimated equilibrium variances beyond sampling variability. 
  
\begin{figure}[htbp]
\centering
\includegraphics[scale=0.35]{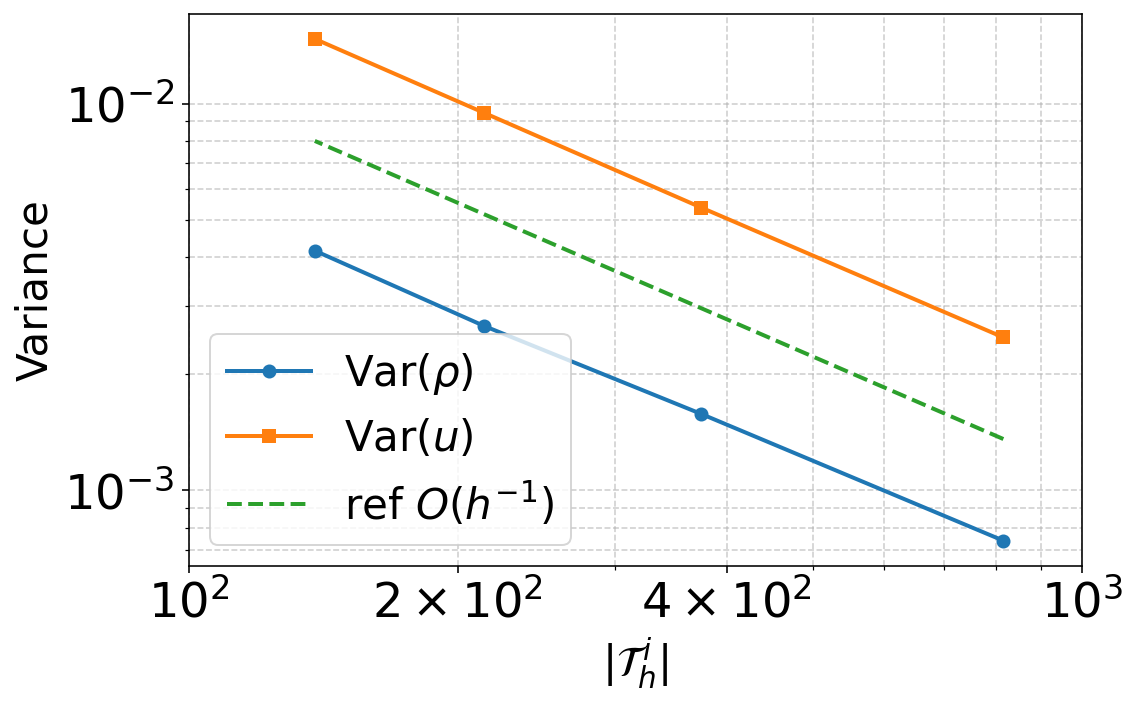}%
\caption{Log--log plot of $\mathrm{Var}(\rho)$ and $\mathrm{Var}(u)$ vs.\ grid spacing $|T_h^i|^{-1}$ for $\Delta t=0.5$, together with a reference line of order $O(|\mathcal T_h^i|^{-1})$.}
\label{fig:var_loglog_dt05}
\end{figure}

Table~\ref{tab:var_dt05_h_2d} summarizes the time-averaged variance estimates
$\langle\widehat{s}_u^{\,2}\rangle$ and $\langle\widehat{s}_\rho^{\,2}\rangle$ over the equilibrium window for a fixed time step $\Delta t\in\{0.2,\,0.4,\,0.8\,1.6\}$. The estimates remain within a few percent of the theoretical predictions across all tested mesh sizes. Moreover, the empirical rates $p_u$ and $p_\rho$ are close to $1$, which is consistent with the scaling $\langle\delta(\cdot)^2\rangle \propto |\mathcal T_h^i|^{-1}$ in two dimensions: when the mesh area is reduced by a factor of two, the element area decreases by a factor of two and the variance increases by the same factor. This trend is also visible in the log--log representation in Figure~\ref{fig:var_loglog_dt05}, where the variance estimates follow the expected $O(|\mathcal T_h^i|^{-1})$ scaling.

\begin{table}[htbp]
\centering
\caption{Two-dimensional example: average sample variances $\langle\widehat{s}_u^{\,2}\rangle$ and $\langle\widehat{s}_\rho^{\,2}\rangle$ for $\Delta t=0.5$, where
$\langle\widehat{s}^{\,2}\rangle$ denotes the mean of $\widehat{s}^{\,2}(t_n)$ over the equilibrium window. The theoretical values are 
$\langle \delta u^2\rangle$ and $\langle \delta\rho^2\rangle$. 
Relative errors are $\big(\widehat{s}^{\,2}-\langle\cdot\rangle\big)/\langle\cdot\rangle\times 100\%$,
and the empirical rate is
$p=\log\!\big(\widehat{s}^{\,2}_{\mathrm{finer}}/\widehat{s}^{\,2}_{\mathrm{coarser}}\big)/\log\!\big(h_{\text{finer}}/h_{\text{coarser}}\big)$.}
\label{tab:var_dt05_h_2d}
\scalebox{0.88}{%
\setlength{\tabcolsep}{5pt}
\begin{tabular}{
  S[table-format=3.1]
  S[table-format=1.6] 
  S[table-format=1.6] 
  S[table-format=1.2]
  S[table-format=1.6] 
  S[table-format=1.6] 
  S[table-format=+1.2]
  S[table-format=1.3]
  S[table-format=1.3]
}
\toprule
& \multicolumn{3}{c}{Velocity} & \multicolumn{3}{c}{Density} & \multicolumn{2}{c}{Rate} \\
\cmidrule(lr){2-4}\cmidrule(lr){5-7}\cmidrule(lr){8-9}
{$|\mathcal{T}^i_h|$} &
{$\langle\hat{s}_u^{\,2}\rangle$} & {$\langle \delta u^2\rangle$} & {Error (\%)} &
{$\langle\widehat{s}_{\rho}^{\,2}\rangle$} & {$\langle \delta\rho^2\rangle$} & {Error (\%)} &
{$p_\rho$} & {$p_u$} \\
\midrule
816.3 & 0.002485 & 0.002450 & +1.43 & 0.000741 & 0.000735 & +0.83 & \multicolumn{1}{c}{---} & \multicolumn{1}{c}{---} \\
374.6 & 0.005384 & 0.005339 & +0.84 & 0.001576 & 0.001602 & -1.61 & 0.969 & 0.992 \\
214.2 & 0.009455 & 0.009339 & +1.25 & 0.002659 & 0.002802 & -5.08 & 0.936 & 1.007 \\
138.4 & 0.014693 & 0.014450 & +1.68 & 0.004164 & 0.004335 & -3.95 & 1.027 & 1.010 \\
\bottomrule
\end{tabular}
}
\end{table}

\begin{figure}[htbp]
\centering
\begin{minipage}{0.48\textwidth}
\centering
\includegraphics[width=\linewidth]{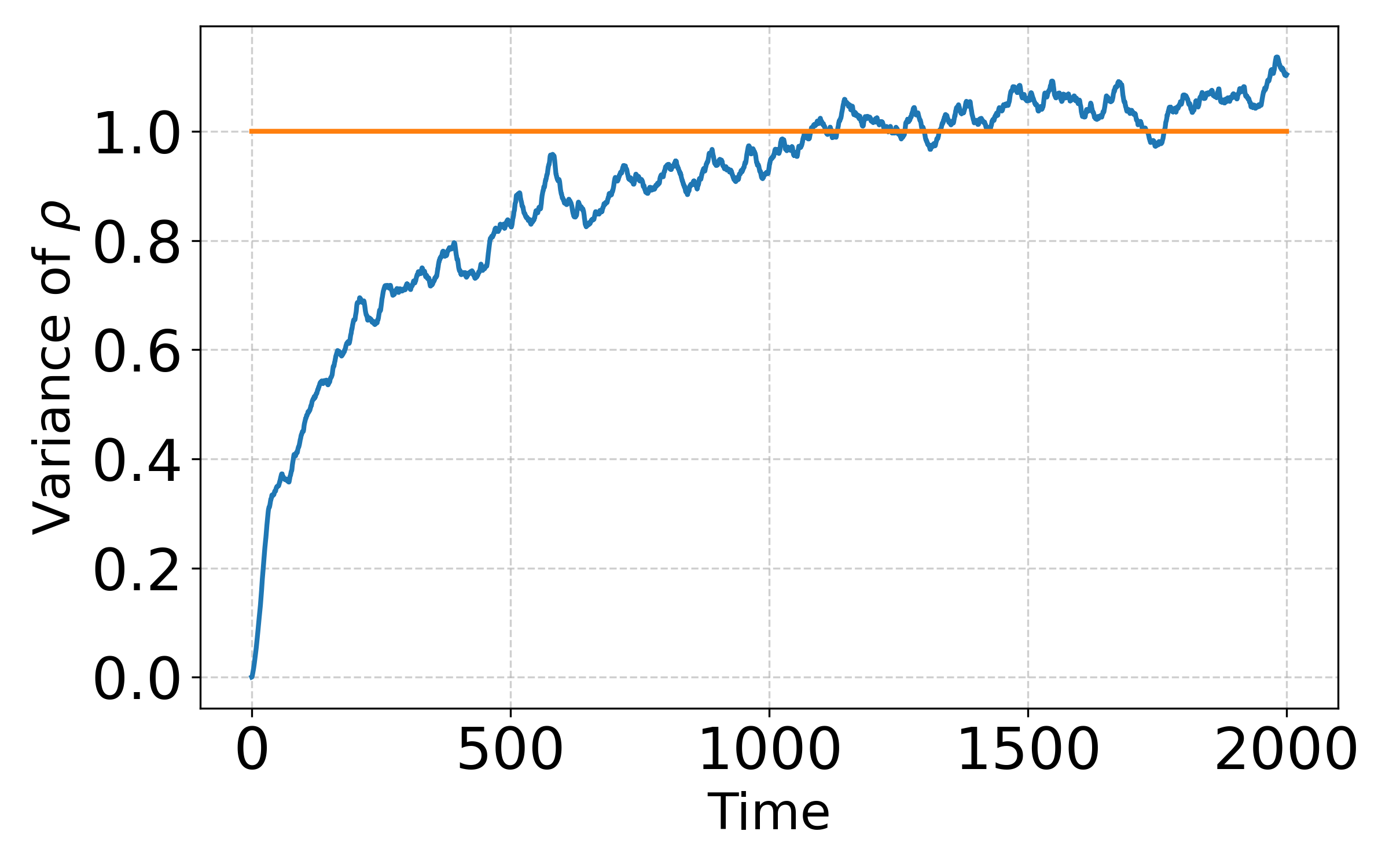}
\end{minipage}\hfill
\begin{minipage}{0.48\textwidth}
\centering
\includegraphics[width=\linewidth]{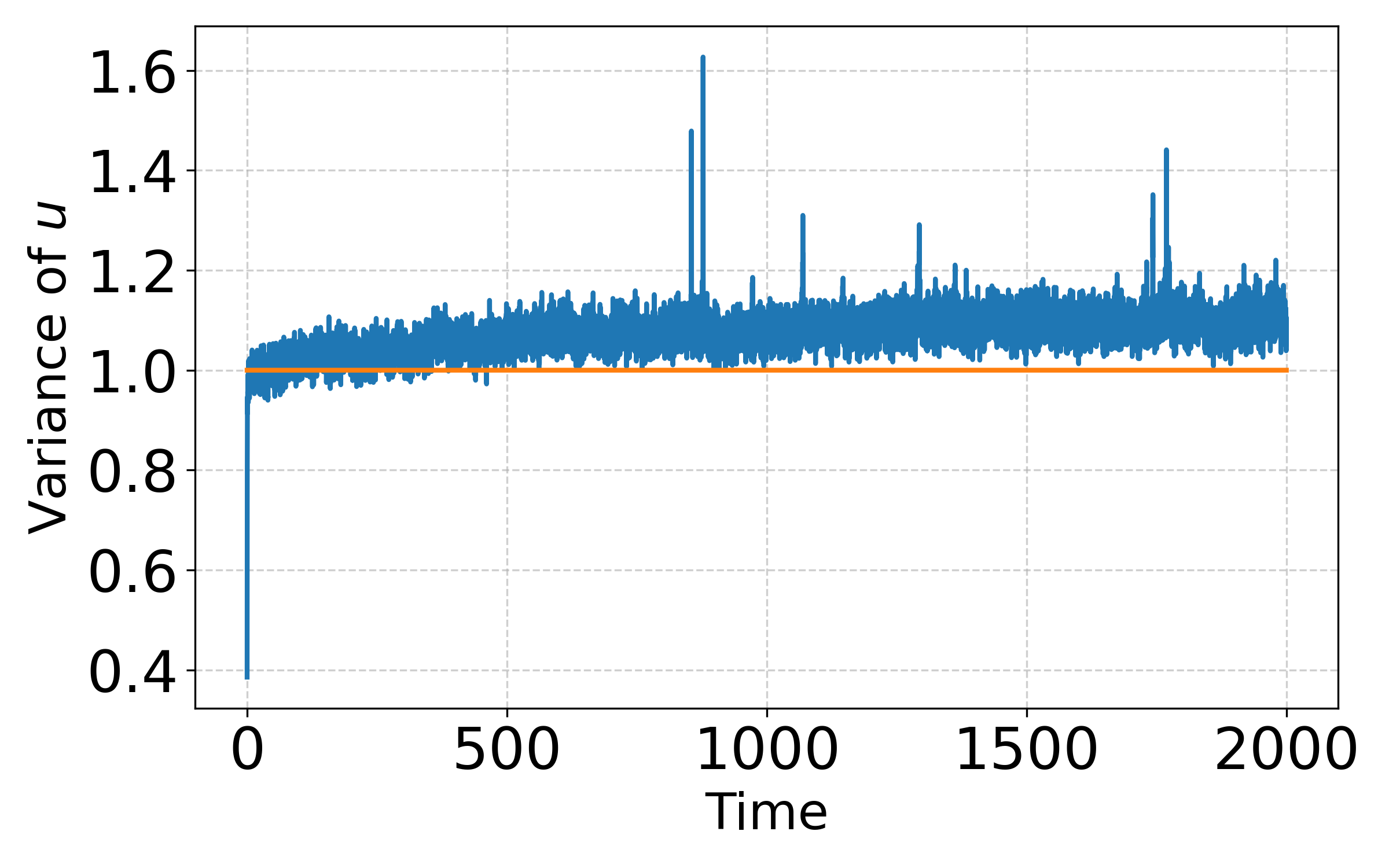}
\end{minipage}

\par\smallskip
{\footnotesize\centering $\Delta t=0.1,\; h=138.4$.\par}

\caption{Two-dimensional example: triangular mesh;time evolution of the normalised sample variances of the density $\rho=0.1$ (left) and velocity $u$ (right) for $\Delta t=0.1$ and $h=138.6$. The horizontal orange line indicates the equilibrium reference level $1$.}
\label{fig:variance_dt_h_2d_small_rho}
\end{figure}

Finally, for small mean densities, the normalised variance estimates deviate from the
linear equilibrium predictions. For example, in Figure~\ref{fig:variance_dt_h_2d_small_rho} with $\langle\rho\rangle=0.1$ the estimated variance levels no longer match
\eqref{eq:var_rho_theory}--\eqref{eq:var_u_theory}, which is consistent with the reduced validity of the linearized equilibrium formulas \cite{de2006hydrodynamic}. 

\begin{figure}[htbp]
\centering

\begin{minipage}{0.48\textwidth}
\centering
\includegraphics[width=\linewidth]{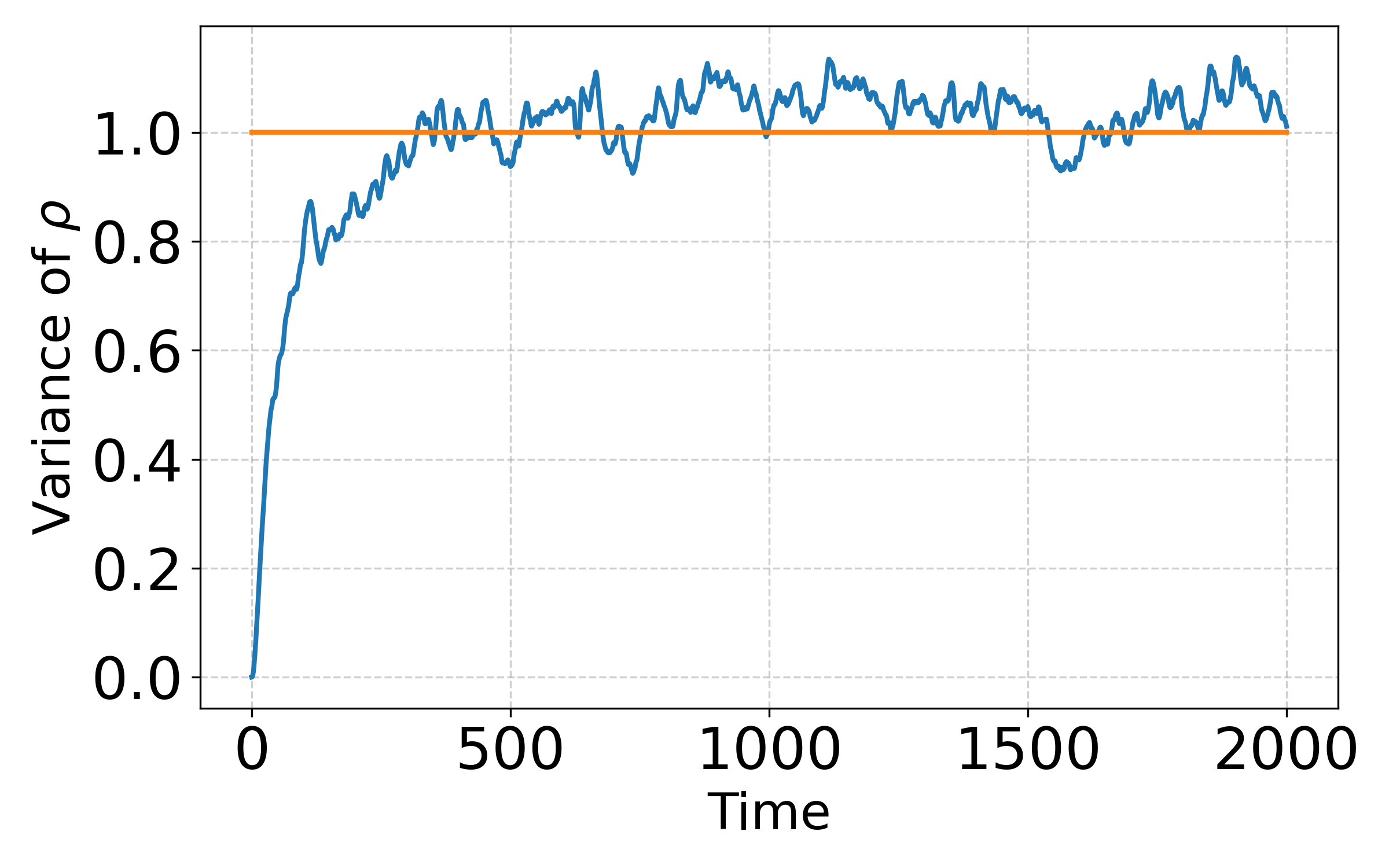}
\end{minipage}\hfill
\begin{minipage}{0.48\textwidth}
\centering
\includegraphics[width=\linewidth]{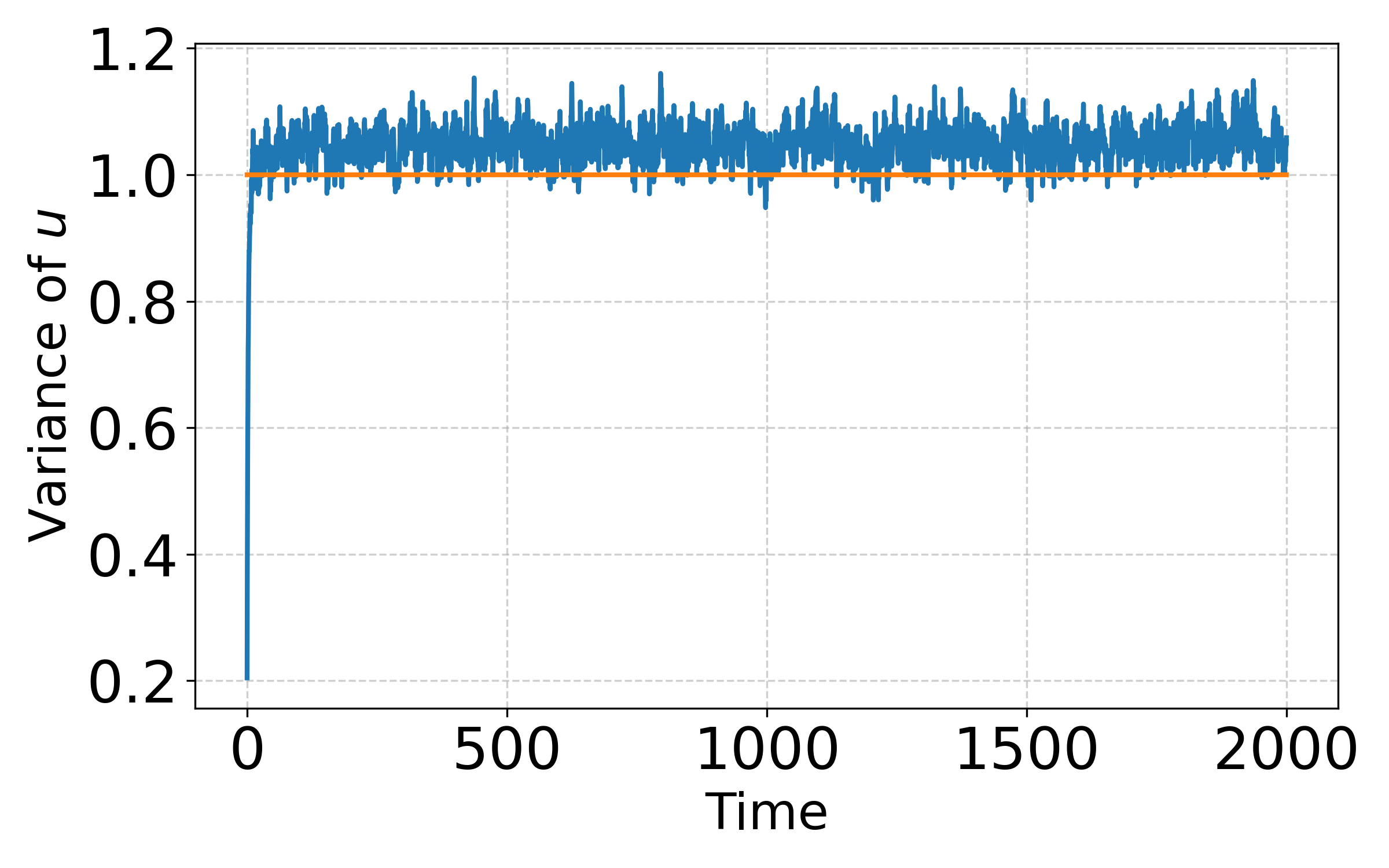}
\end{minipage}

\par\smallskip
{\footnotesize\centering $\Delta t=0.4,\; \mathcal{T}^i_h=6010.5$.\par}

\vspace{0.8em}

\begin{minipage}{0.48\textwidth}
\centering
\includegraphics[width=\linewidth]{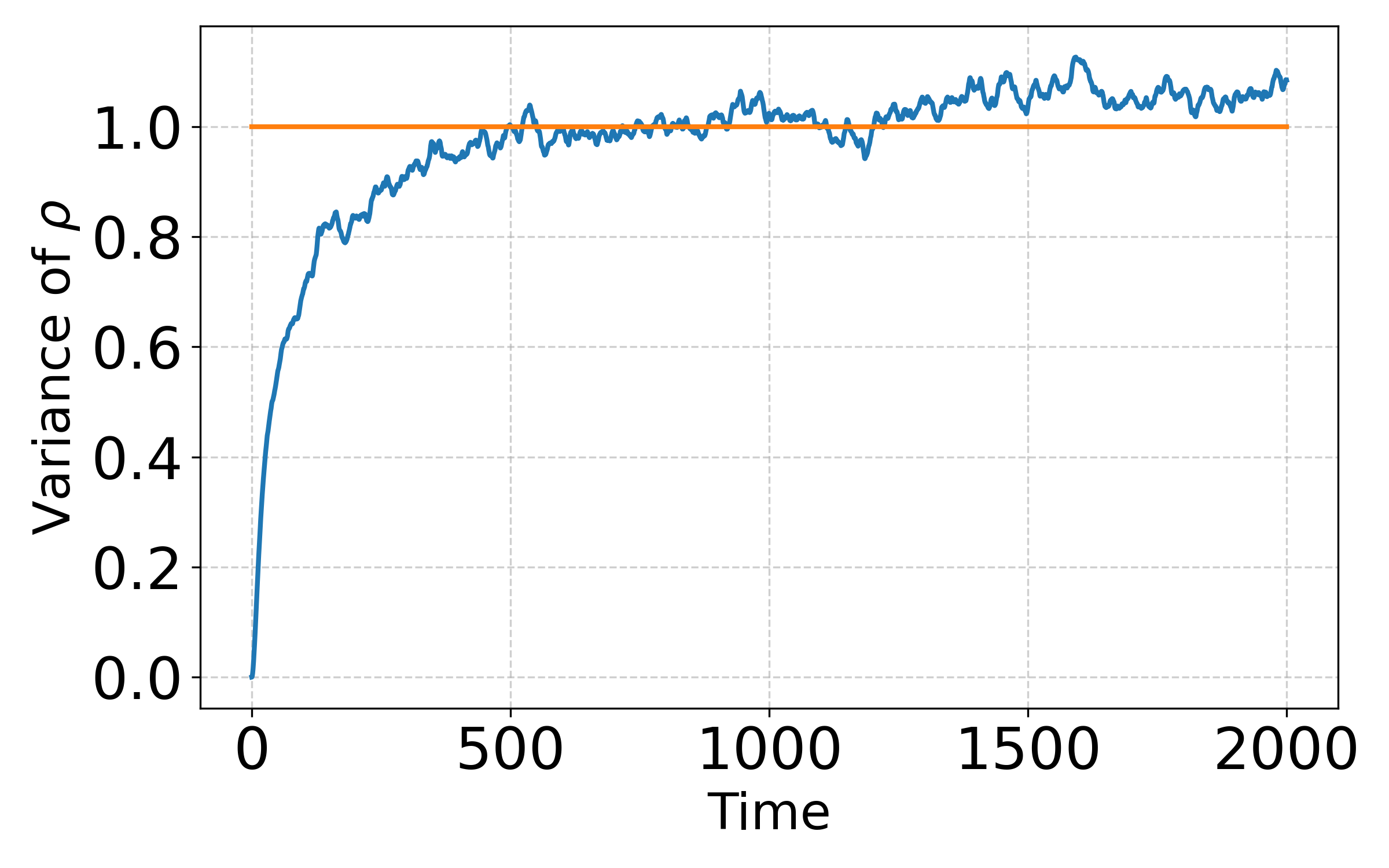}
\end{minipage}\hfill
\begin{minipage}{0.48\textwidth}
\centering
\includegraphics[width=\linewidth]{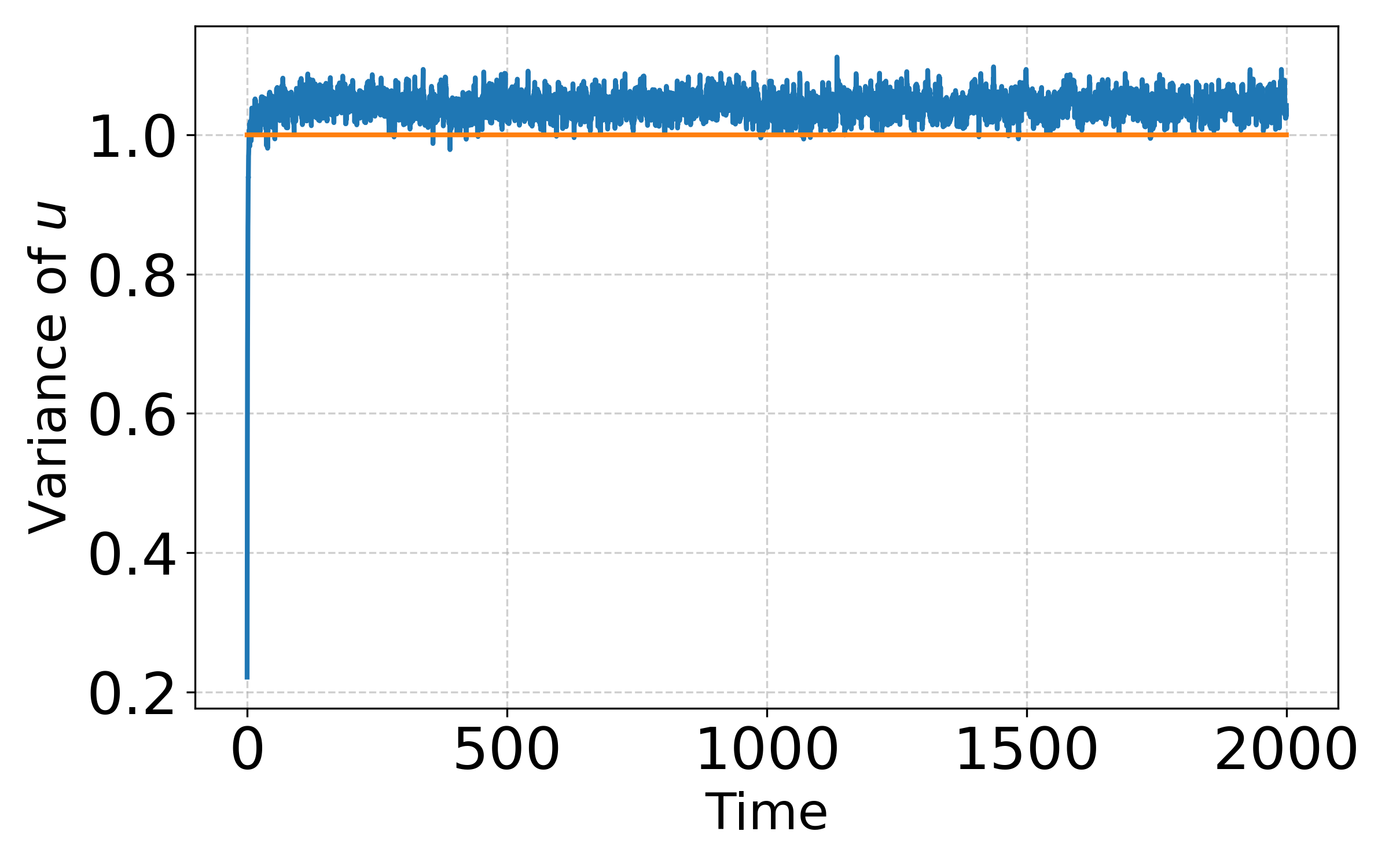}
\end{minipage}

\par\smallskip
{\footnotesize\centering $\Delta t=0.2,\; \mathcal{T}^i_h=1953.1$.\par}

\vspace{0.8em}

\begin{minipage}{0.48\textwidth}
\centering
\includegraphics[width=\linewidth]{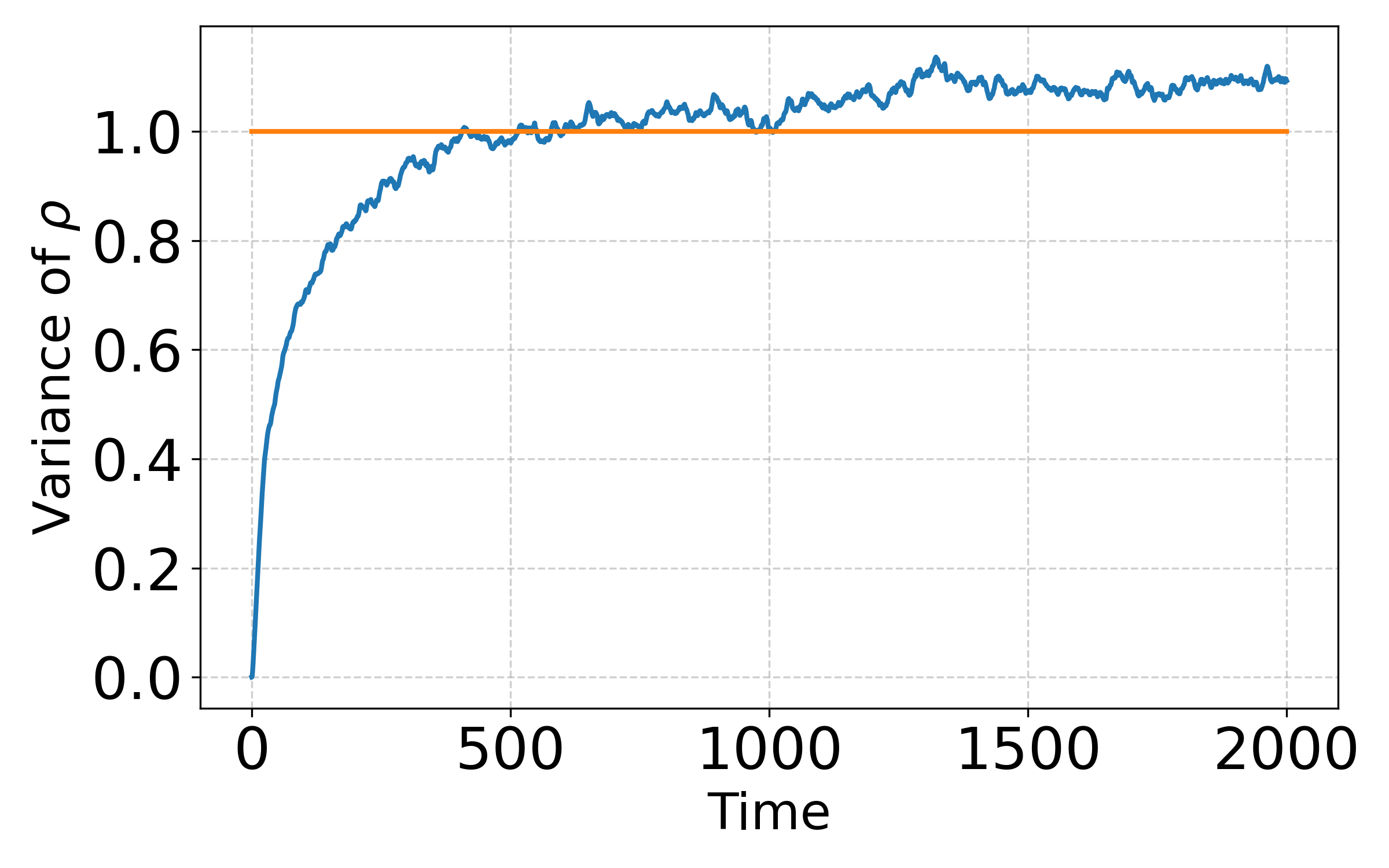}
\end{minipage}\hfill
\begin{minipage}{0.48\textwidth}
\centering
\includegraphics[width=\linewidth]{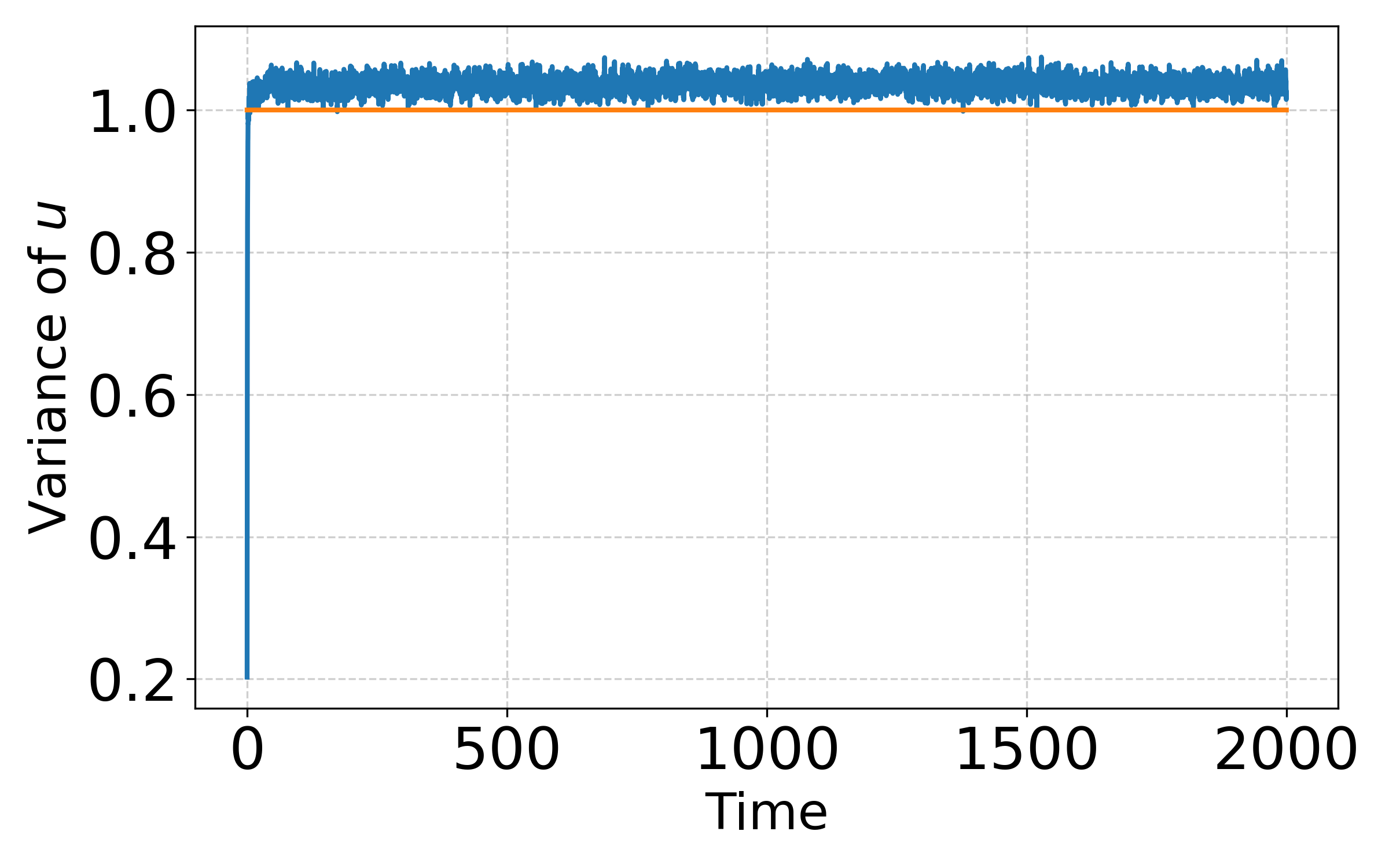}
\end{minipage}

\par\smallskip
{\footnotesize\centering $\Delta t=0.1,\; \mathcal{T}^i_h=863.8$.\par}

\caption{Three-dimensional example: time evolution of the normalised sample variances of the density $\rho$ (left) and velocity $u$ (right) for $\Delta t\in\{0.4,\,0.2,\,0.1\}$ and $\mathcal{T}^i_h\in\{6010.5,1953.1,863.8\}$ (top to bottom). The horizontal orange line indicates the equilibrium reference level $1$.}

\label{fig:variance_dt_h_3d}
\end{figure}

We next repeat the equilibrium test in three dimensions on the cubic domain
$\Omega=(0,200)^3$.
Figure~\ref{fig:variance_dt_h_3d} shows the normalised variance estimates for decreasing mesh size and time step.
Once equilibrium is reached, both estimates fluctuate around their reference levels; the density variance remains centered at $1$, while the velocity variance
stays slightly above unity. In contrast to the one-dimensional case, the velocity variance reaches its steady level almost immediately, essentially independent of the mesh size and the time step.

\begin{figure}[htbp]
\centering
\begin{minipage}{0.48\textwidth}
\centering
\includegraphics[width=\linewidth]{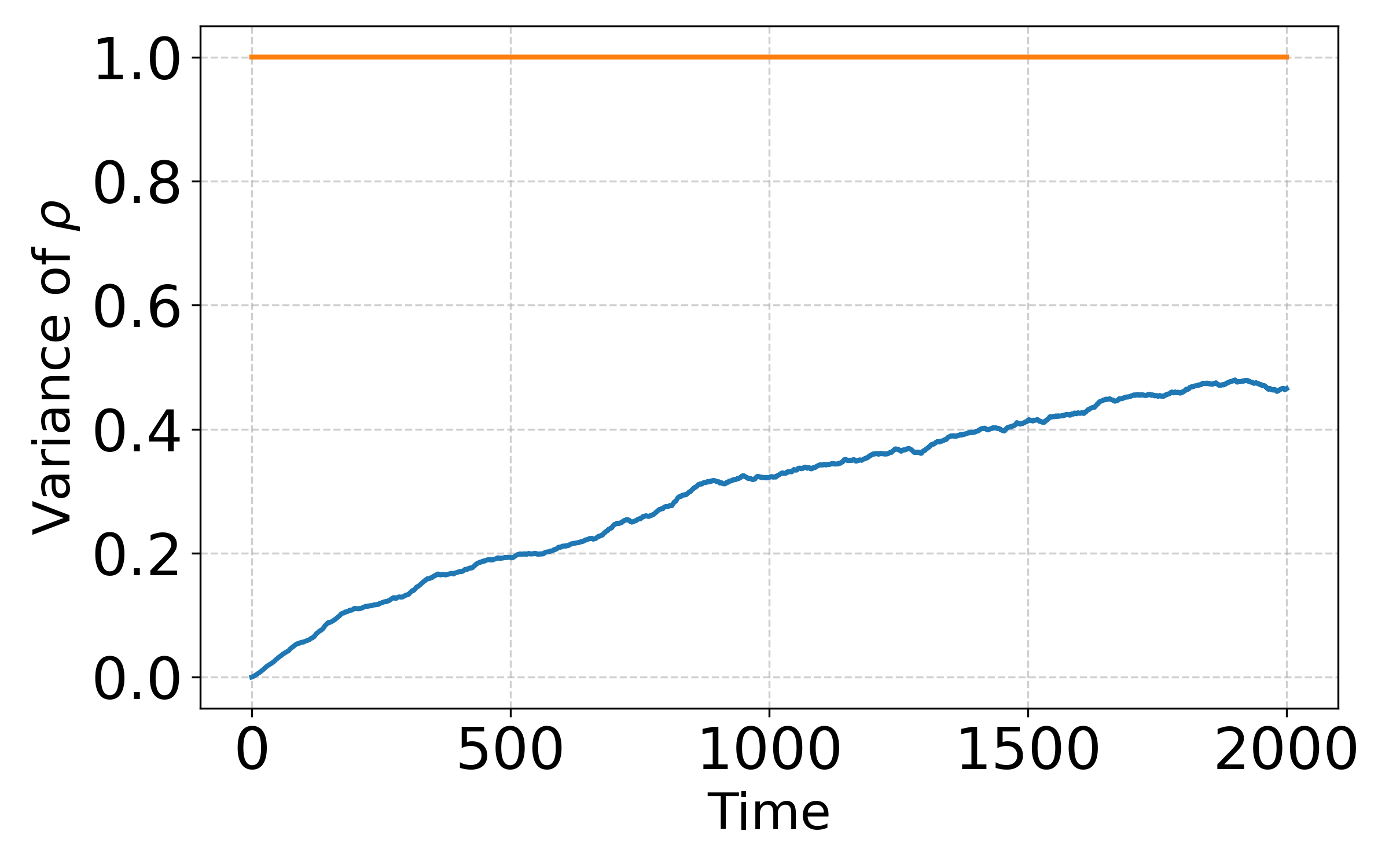}
\end{minipage}\hfill
\begin{minipage}{0.48\textwidth}
\centering
\includegraphics[width=\linewidth]{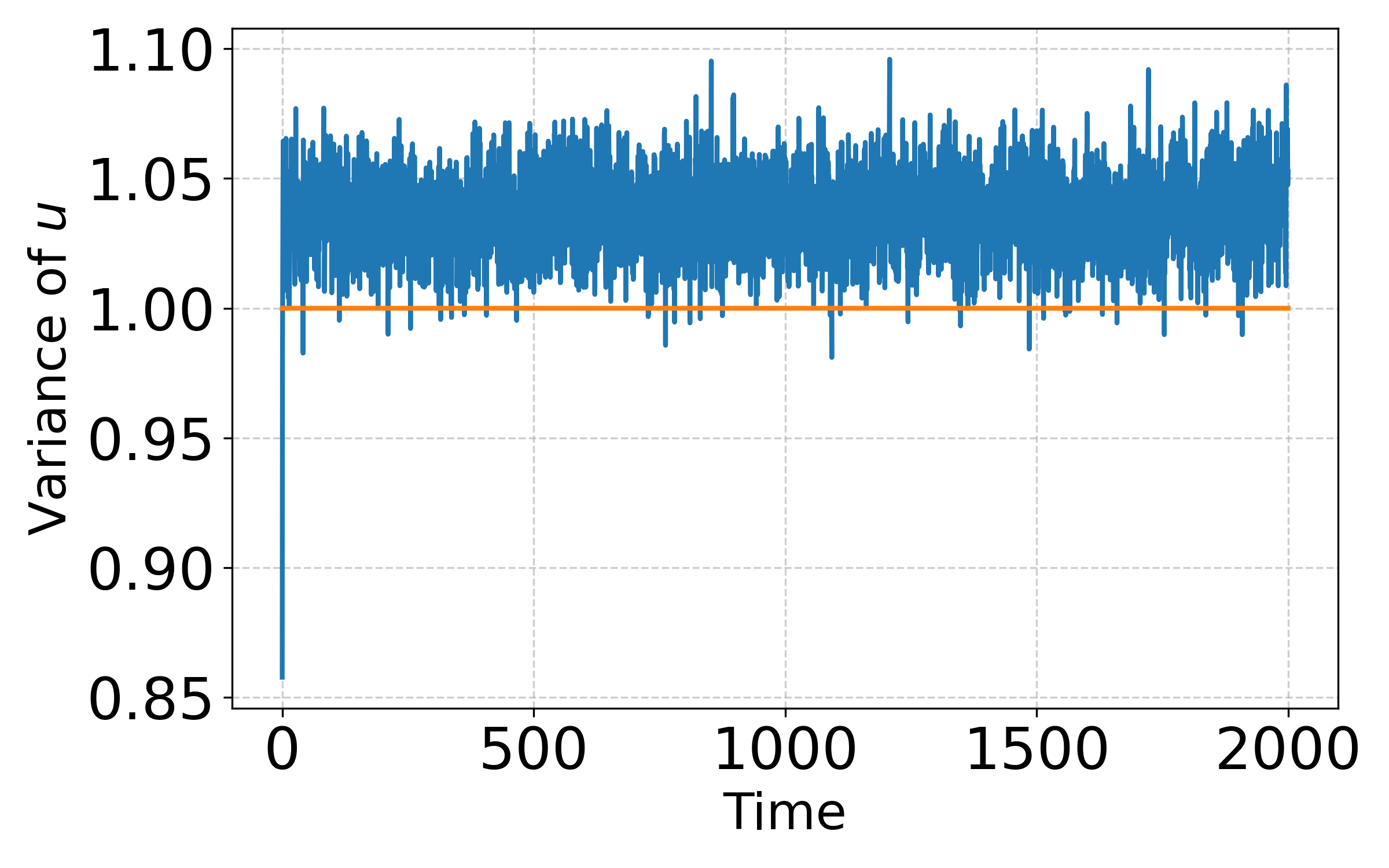}
\end{minipage}

\par\smallskip
{\footnotesize\centering $\Delta t=0.5,\; |\mathcal{T}^i_h|=52734.3$.\par}

\caption{Three-dimensional example: time evolution of the normalised sample variances of the density $\rho=0.01$ (left) and velocity $u$ (right) for $\Delta t=0.5$ and $|\mathcal{T}^i_h|=52734.3$. The horizontal orange line indicates the equilibrium reference level $1$.}
\label{fig:3dvariance_dt_h_2d_small_rho}
\end{figure}

Table~\ref{tab:var_3d} summarizes the time-averaged variance estimates
 computed over the equilibrium window for a sequence of runs in which the time step is halved. Across all tested mesh sizes, the estimates remain within a few percent of the theoretical predictions. The corresponding empirical refinement rates $p_u$ and $p_\rho$ are close to $1$, in agreement with the expected scaling. Figure~\ref{fig:3dvariance_dt_h_2d_small_rho} shows that, for sufficiently small mean density
$\langle\rho\rangle=10^{-2}$, the normalised variance estimates deviate from the linear
equilibrium values as in the
two-dimensional case.

\begin{table}[htbp]
\centering
\caption{Three-dimensional example: average sample variances $\langle\widehat{s}_u^{\,2}\rangle$ and $\langle\widehat{s}_\rho^{\,2}\rangle$, where
$\langle\widehat{s}^{\,2}\rangle$ denotes the mean of $\widehat{s}^{\,2}(t_n)$ over the equilibrium window. The theoretical values are 
$\langle \delta u^2\rangle$ and $\langle \delta\rho^2\rangle$. 
Relative errors are $\big(\widehat{s}^{\,2}-\langle\cdot\rangle\big)/\langle\cdot\rangle\times 100\%$,
and the empirical rate is
$p=\log\!\big(\widehat{s}^{\,2}_{\mathrm{finer}}/\widehat{s}^{\,2}_{\mathrm{coarser}}\big)/\log\!\big(|\mathcal T^i_{h,\mathrm{coarser}}|/|\mathcal T^i_{h,\mathrm{finer}}|\big)$.}
\label{tab:var_3d}

\scalebox{0.88}{%
\setlength{\tabcolsep}{5pt}
\begin{tabular}{
  S[table-format=5.1]
  S[table-format=1.6] 
  S[table-format=1.6] 
  S[table-format=1.2]
  S[table-format=1.6] 
  S[table-format=1.6] 
  S[table-format=+1.3]
  S[table-format=1.3]
  S[table-format=1.3]
}
\toprule
& \multicolumn{3}{c}{Velocity} & \multicolumn{3}{c}{Density} & \multicolumn{2}{c}{Rate} \\
\cmidrule(lr){2-4}\cmidrule(lr){5-7}\cmidrule(lr){8-9}
{$|\mathcal{T}^i_h|$} &
{$\langle\hat{s}_u^{\,2}\rangle$} & {$\langle \delta u^2\rangle$} & {Error (\%)} &
{$\langle\widehat{s}_{\rho}^{\,2}\rangle$} & {$\langle \delta\rho^2\rangle$} & {Error (\%)} &
{$p_\rho$} & {$p_u$} \\
\midrule
37037.0 & 0.000056 & 0.000054 & +3.98 & 0.000017 & 0.000016 & +4.77 & \multicolumn{1}{c}{---} & \multicolumn{1}{c}{---} \\
6010.5  & 0.000349 & 0.000333 & +4.81 & 0.000101 & 0.000100 & +0.75 & 0.978 & 1.004 \\
1953.1  & 0.001068 & 0.001024 & +4.26 & 0.000305 & 0.000307 & -0.62 & 0.988 & 0.995 \\
863.8   & 0.002399 & 0.002315 & +3.61 & 0.000702 & 0.000695 & +1.04 & 1.020 & 0.992 \\
\bottomrule
\end{tabular}
}
\end{table}

\subsection{Additional comments on the numerical method}

The two- and three-dimensional cases require substantially more computational time and memory due to the larger number of degrees of freedom. For the same coarse-graining, however, the effective stochastic forcing is stronger in one dimension, making the density more prone to becoming negative, especially at small mean values. In this work, we do not explicitly enforce the constraint $\rho\ge 0$ at the discrete level; nevertheless, we can reproduce the results of \cite{MagalettiGalloPerezCarrilloKalliadasis}, which were obtained with a positivity-preserving finite-difference method. Empirically, the lowest mean densities we can treat reliably are $\rho\approx 1$ in one dimension, $\rho\approx 10^{-1}$ in two dimensions, and $\rho\approx 10^{-2}$ in three dimensions.

While \cite{RussoPerezDuranYatsyshinCarrilloKalliadasis} enforces positivity through a Brownian-bridge-type construction in a single-equation setting,
extending this idea to our work is not straightforward. Here, $\rho$ is connected to the continuity equation and is coupled to the stochastic velocity update. Enforcing $\rho\ge 0$ via a Brownian-bridge correction would require an extension to the construction in Ref. \cite{RussoPerezDuranYatsyshinCarrilloKalliadasis} by implementing a technique corresponding to the integral of the Brownian bridge instead. In doing so, one must treat the density and velocity fields together as a multidimensional stochastic variable to preserve the Markov condition, and apply the bridge methodology to this joint process. Additionally, due to the nonlinearity of the fluctuating Navier-Stokes equations the stochastic process is not Gaussian, which adds further complexities to the bridge construction. Thus, it may only be feasible to implement the bridge for the equations of linearised fluctuating hydrodynamics. Due to these intricacies, designing a rigorously positivity-preserving variant for the coupled system is left for future work.

Finally, we remark that in all dimensions tested the total mass is conserved, despite using a continuous Galerkin discretisation. This point is important because enforcing additional monotonicity or positivity properties by introducing upwinding or other non-symmetric stabilisations typically modifies the discrete dissipation and can alter the fluctuation--dissipation balance. A possible compromise is to use a hybrid discretisation that retains a centered treatment of the stochastic fluctuations while applying limited upwinding only where needed for robustness, as suggested in \cite{MagalettiGalloPerezCarrilloKalliadasis}.

\section{Conclusions}

In this work, we have developed and validated a structure-preserving finite element framework for simulating thermal fluctuations in compressible fluids governed by the isothermal fluctuating Navier–Stokes equations. Our methodology ensures discrete consistency with the fluctuation–dissipation theorem by constructing a stochastic forcing term whose covariance explicitly matches the discrete viscous dissipation operator. To overcome the well-known issue of artificial long-range correlations in classical finite element discretisations, we introduced a nodal quadrature rule, which effectively localizes fluctuations and yields physically realistic equilibrium statistics.

The numerical scheme was thoroughly verified across one, two, and three spatial dimensions. Results confirm that the method correctly reproduces theoretical variance scaling with mesh refinement, maintains mass conservation, and captures expected equilibrium distributions across a range of discretisation parameters. 

While the method performs robustly for moderate fluctuation amplitudes, future work will focus on enforcing positivity constraints for stronger fluctuations and extending the approach to multi-component and non-isothermal systems. The flexibility of the finite element method also opens the door to applications involving complex geometries and adaptive mesh refinement, providing a powerful tool for multiscale simulation of fluctuating hydrodynamics in realistic settings.

\clearpage

\appendix
\section{Stochastic analysis of the fluctuating stress in the weak formulation} \label{Appendix A}

Define the stochastic forcing induced by the fluctuating stress as the linear functional
\begin{equation}\label{eq:Fvw_def}
F(\mathbf v,t)
:= \int_\Omega \big(\nabla\!\cdot \widetilde{\boldsymbol\tau}(x,t)\big)\cdot \mathbf v(x)\,dx
= -\int_\Omega \widetilde{\boldsymbol\tau}(x,t) : \nabla \mathbf v(x)\,dx,
\end{equation}
 here the second identity follows by integration by parts under boundary conditions
for which the boundary term vanishes.

Assume the Landau--Lifshitz statistics
\begin{flalign}
\label{eq:tau_cov_app_vw}
\langle \widetilde{\tau}_{ij}(x,t)\rangle &= 0, \nonumber\\
\langle \widetilde{\tau}_{ij}(x,t)\,\widetilde{\tau}_{pq}(y,s)\rangle
&= 2\mu k_BT \,
\Big(\delta_{ip}\delta_{jq}+\delta_{iq}\delta_{jp}-\tfrac{2}{d}\delta_{ij}\delta_{pq}\Big)\,
\delta(x-y)\,\delta(t-s).
\end{flalign}
Then, using \eqref{eq:Fvw_def}--\eqref{eq:tau_cov_app_vw} and the identity
$\int_\Omega\!\!\int_\Omega \delta(x-y)\,g(x,y)\,dx\,dy=\int_\Omega g(x,x)\,dx$, we obtain
\begin{align}
\left\langle F(\mathbf v,t)\,F(\mathbf w,s)\right\rangle
&= \int_\Omega\!\!\int_\Omega
\langle \widetilde{\tau}_{ij}(x,t)\,\widetilde{\tau}_{pq}(y,s)\rangle\,
\partial_j v_i(x)\,\partial_q w_p(y)\,dx\,dy \notag\\
&= 2\mu k_BT\,\delta(t-s)\int_\Omega
\Big(\delta_{ip}\delta_{jq}+\delta_{iq}\delta_{jp}-\tfrac{2}{d}\delta_{ij}\delta_{pq}\Big)\,
\partial_j v_i(x)\,\partial_q w_p(x)\,dx.
\label{eq:Fcov_index_vw}
\end{align}
Equivalently, in tensor notation,
\begin{equation}\label{eq:Fcov_compact_vw}
\left\langle F(\mathbf v,t)\,F(\mathbf w,s)\right\rangle
= 2\mu k_BT\,\delta(t-s)\int_\Omega
\Big(\nabla \mathbf v + (\nabla \mathbf v)^{\!\top}
-\tfrac{2}{d}(\nabla\!\cdot \mathbf v)\,\mathbf I\Big)
:
\nabla \mathbf w
\,dx,
\end{equation}
which yields \eqref{eq:weak-covariance} with the viscous bilinear form
$a(\cdot,\cdot)$ defined in \eqref{eq:viscous-bilinear-form}.

\section{One-Dimensional Example: Mass Matrix and Mass \\ Lumping} \label{Appendix B}

Let $\{\phi_i\}_{i=0}^N$ be the standard one-dimensional $P_1$ nodal basis functions on the mesh
\[
a=x_0<x_1<\cdots<x_N=b.
\]
For an interior node $x_i$, the basis function $\phi_i$ is given by
\[
\phi_i(x)=
\begin{cases}
\dfrac{x-x_{i-1}}{x_i-x_{i-1}}, & x\in[x_{i-1},x_i], \\[1ex]
\dfrac{x_{i+1}-x}{x_{i+1}-x_i}, & x\in[x_i,x_{i+1}], \\[1ex]
0, & \text{otherwise.}
\end{cases}
\]
Therefore, $\phi_i$ is supported on $[x_{i-1},x_{i+1}]$ and overlaps only with its immediate neighbors $\phi_{i-1}$ and $\phi_{i+1}$.

\begin{figure}[htbp]
\centering

\begin{minipage}{0.48\textwidth}
    \centering
    \includegraphics[width=\linewidth]{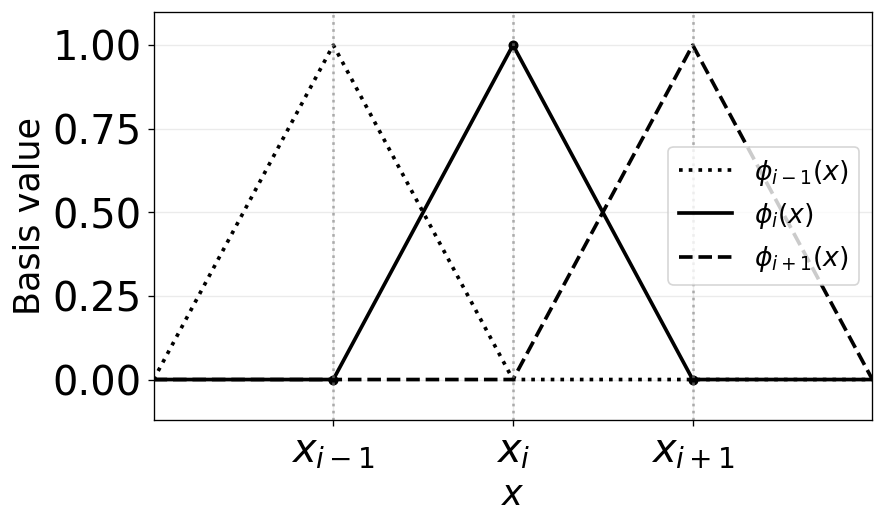}
\end{minipage}\hfill
\begin{minipage}{0.48\textwidth}
    \centering
    \includegraphics[width=\linewidth]{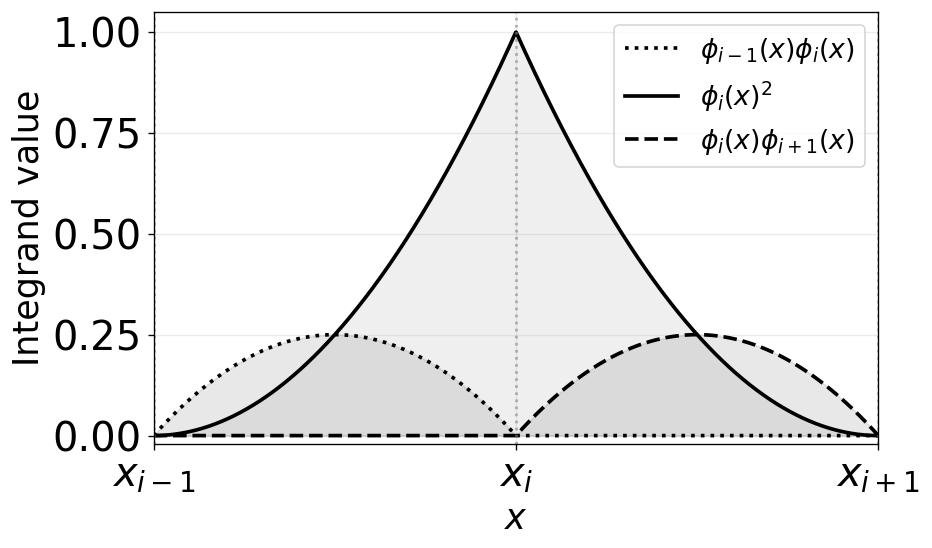}
\end{minipage}

\caption{Illustration of the one-dimensional linear finite element basis and the associated mass-matrix integrands. Left: local hat basis functions $\phi_{i-1}$, $\phi_i$, and $\phi_{i+1}$. Right: the integrands $\phi_{i-1}\phi_i$, $\phi_i^2$, and $\phi_i\phi_{i+1}$, whose integrals define the mass matrix entries.}
\label{fig:mass_matrix_basis_integrands}
\end{figure}

The mass matrix is defined by
$$
M_{ij}=\int_\Omega \phi_i(x)\phi_j(x)\,dx.
$$
Because of the local support of the basis functions, one has $\phi_i\phi_j=0$ for $|i-j|>1$. Therefore, only the entries $M_{i,i-1}$, $M_{ii}$, and $M_{i,i+1}$ are nonzero, so the consistent mass matrix is tridiagonal. In particular, the $i$th component of the mass term is given by
$$
(Mu)_i
=
u_{i-1}\int_\Omega \phi_i\phi_{i-1}\,dx
+
u_i\int_\Omega \phi_i^2\,dx
+
u_{i+1}\int_\Omega \phi_i\phi_{i+1}\,dx.
$$
The corresponding integrands are shown in the right panel of Figure~\ref{fig:mass_matrix_basis_integrands} along with the basis function on the left panel.
Using nodal quadrature on each element,
$$
\int_{x_i}^{x_{i+1}} f(x)\,dx
\approx
\dfrac{h}{2}\bigl(f(x_i)+f(x_{i+1})\bigr),
$$
the integrands corresponding to the off-diagonal entries are zero at the quadrature points, since
$$
\phi_i(x_j)\phi_k(x_j)=0, \qquad i\neq k.
$$
Therefore, the quadrature approximation of the off-diagonal mass entries vanishes, yielding a diagonal lumped mass matrix. Consequently, the original mass term couples the neighboring degrees of freedom $u_{i-1}$, $u_i$, and $u_{i+1}$, whereas the lumped mass term depends only on $u_i$.

\section*{Acknowledgements}
The authors want to thank Miguel A. Durán-Olivencia for valuable discussions.

\bibliographystyle{siamplain}
\bibliography{references}

\end{document}

%% file: macros.tex
\def\frac#1#2{{\textstyle{#1\over#2}}}

\def\bi{\begin{itemize}}
\def\ei{\end{itemize}}
\def\bd{\begin{description}}
\def\ed{\end{description}}
\def\ben{\begin{enumerate}}
\def\een{\end{enumerate}}
\def\bv{\begin{verbatim}}
\def\ev\end{verbatim}

\def\2to{{\ {\buildrel 2\over \longrightarrow}\ }}

\def\I1ton{{$I_1,\ldots,I_n$}}
\def\X1ton{{$X_1,\ldots,X_n$}}
\def\Y1ton{{$Y_1,\ldots,Y_n$}}
\def\Z1ton{{$Z_1,\ldots,Z_n$}}
\def\R1ton{{$R_1,\ldots,R_n$}}
\def\e1ton{{$e_1,\ldots,e_n$}}
\def\t1ton{{$t_1,\ldots,t_n$}}
\def\x1ton{{$x_1,\ldots,x_n$}}
\def\y1ton{{$y_1,\ldots,y_n$}}
\def\z1ton{{$z_1,\ldots,z_n$}}
\def\tt#1{{\texttt{#1}}}